\documentclass[aps,prc,superscriptaddress,twocolumn]{revtex4-1}
\usepackage{graphicx}
\usepackage{amssymb,amsmath}
\usepackage{color}
\usepackage{lineno}
\usepackage{tabularx}
\usepackage{placeins}

\begin{document}

\title{Measurement of Unpolarized and Polarized Cross Sections for Deeply Virtual Compton Scattering on the Proton at Jefferson Laboratory with CLAS}
\
\newcommand*{\ANL}{Argonne National Laboratory, Argonne, Illinois 60439}
\newcommand*{\ANLindex}{1}
\affiliation{\ANL}
\newcommand*{\ASU}{Arizona State University, Tempe, Arizona 85287-1504}
\newcommand*{\ASUindex}{2}
\affiliation{\ASU}
\newcommand*{\CSUDH}{California State University, Dominguez Hills, Carson, CA 90747}
\newcommand*{\CSUDHindex}{3}
\affiliation{\CSUDH}
\newcommand*{\CANISIUS}{Canisius College, Buffalo, NY}
\newcommand*{\CANISIUSindex}{4}
\affiliation{\CANISIUS}
\newcommand*{\CMU}{Carnegie Mellon University, Pittsburgh, Pennsylvania 15213}
\newcommand*{\CMUindex}{5}
\affiliation{\CMU}
\newcommand*{\CUA}{Catholic University of America, Washington, D.C. 20064}
\newcommand*{\CUAindex}{6}
\affiliation{\CUA}
\newcommand*{\SACLAY}{IRFU, CEA, Universit\'e Paris-Saclay, F-91191 Gif-sur-Yvette, France}
\newcommand*{\SACLAYindex}{7}
\affiliation{\SACLAY}
\newcommand*{\CNU}{Christopher Newport University, Newport News, Virginia 23606}
\newcommand*{\CNUindex}{8}
\affiliation{\CNU}
\newcommand*{\UCONN}{University of Connecticut, Storrs, Connecticut 06269}
\newcommand*{\UCONNindex}{9}
\affiliation{\UCONN}
\newcommand*{\DUKE}{Duke University, Durham, North Carolina 27708-0305}
\newcommand*{\DUKEindex}{10}
\affiliation{\DUKE}
\newcommand*{\FU}{Fairfield University, Fairfield CT 06824}
\newcommand*{\FUindex}{11}
\affiliation{\FU}
\newcommand*{\FERRARAU}{Universita' di Ferrara , 44121 Ferrara, Italy}
\newcommand*{\FERRARAUindex}{12}
\affiliation{\FERRARAU}
\newcommand*{\FIU}{Florida International University, Miami, Florida 33199}
\newcommand*{\FIUindex}{13}
\affiliation{\FIU}
\newcommand*{\FSU}{Florida State University, Tallahassee, Florida 32306}
\newcommand*{\FSUindex}{14}
\affiliation{\FSU}
\newcommand*{\Genova}{Universit$\grave{a}$ di Genova, 16146 Genova, Italy}
\newcommand*{\Genovaindex}{15}
\affiliation{\Genova}
\newcommand*{\GWUI}{The George Washington University, Washington, DC 20052}
\newcommand*{\GWUIindex}{16}
\affiliation{\GWUI}
\newcommand*{\IABF} {Imam Abdulrahman Bin Faisal University, Industrial Jubail 31961, Saudi Arabia}
\newcommand*{\IABFindex}{17}
\affiliation{\IABF}
\newcommand*{\ISU}{Idaho State University, Pocatello, Idaho 83209}
\newcommand*{\ISUindex}{17}
\affiliation{\ISU}
\newcommand*{\INFNFE}{INFN, Sezione di Ferrara, 44100 Ferrara, Italy}
\newcommand*{\INFNFEindex}{18}
\affiliation{\INFNFE}
\newcommand*{\INFNFR}{INFN, Laboratori Nazionali di Frascati, 00044 Frascati, Italy}
\newcommand*{\INFNFRindex}{19}
\affiliation{\INFNFR}
\newcommand*{\INFNGE}{INFN, Sezione di Genova, 16146 Genova, Italy}
\newcommand*{\INFNGEindex}{20}
\affiliation{\INFNGE}
\newcommand*{\INFNRO}{INFN, Sezione di Roma Tor Vergata, 00133 Rome, Italy}
\newcommand*{\INFNROindex}{21}
\affiliation{\INFNRO}
\newcommand*{\INFNTUR}{INFN, Sezione di Torino, 10125 Torino, Italy}
\newcommand*{\INFNTURindex}{22}
\affiliation{\INFNTUR}
\newcommand*{\ORSAY}{Institut de Physique Nucl\'eaire, CNRS/IN2P3 and Universit\'e Paris Sud, Orsay, France}
\newcommand*{\ORSAYindex}{23}
\affiliation{\ORSAY}
\newcommand*{\ITEP}{Institute of Theoretical and Experimental Physics, Moscow, 117259, Russia}
\newcommand*{\ITEPindex}{24}
\affiliation{\ITEP}
\newcommand*{\JMU}{James Madison University, Harrisonburg, Virginia 22807}
\newcommand*{\JMUindex}{25}
\affiliation{\JMU}
\newcommand*{\KNU}{Kyungpook National University, Daegu 41566, Republic of Korea}
\newcommand*{\KNUindex}{26}
\affiliation{\KNU}
\newcommand*{\MISS}{Mississippi State University, Mississippi State, MS 39762-5167}
\newcommand*{\MISSindex}{27}
\affiliation{\MISS}
\newcommand*{\UNH}{University of New Hampshire, Durham, New Hampshire 03824-3568}
\newcommand*{\UNHindex}{28}
\affiliation{\UNH}
\newcommand*{\NSU}{Norfolk State University, Norfolk, Virginia 23504}
\newcommand*{\NSUindex}{29}
\affiliation{\NSU}
\newcommand*{\OHIOU}{Ohio University, Athens, Ohio  45701}
\newcommand*{\OHIOUindex}{30}
\affiliation{\OHIOU}
\newcommand*{\ODU}{Old Dominion University, Norfolk, Virginia 23529}
\newcommand*{\ODUindex}{31}
\affiliation{\ODU}
\newcommand*{\URICH}{University of Richmond, Richmond, Virginia 23173}
\newcommand*{\URICHindex}{32}
\affiliation{\URICH}
\newcommand*{\ROMAII}{Universita' di Roma Tor Vergata, 00133 Rome Italy}
\newcommand*{\ROMAIIindex}{33}
\affiliation{\ROMAII}
\newcommand*{\MSU}{Skobeltsyn Institute of Nuclear Physics, Lomonosov Moscow State University, 119234 Moscow, Russia}
\newcommand*{\RPI}{Rensselaer Polytechnic Institute, Troy, New York 12180}
\newcommand*{\RPIindex}{1}
\affiliation{\RPI}
\newcommand*{\MSUindex}{34}
\affiliation{\MSU}
\newcommand*{\SCAROLINA}{University of South Carolina, Columbia, South Carolina 29208}
\newcommand*{\SCAROLINAindex}{35}
\affiliation{\SCAROLINA}
\newcommand*{\TEMPLE}{Temple University,  Philadelphia, PA 19122 }
\newcommand*{\TEMPLEindex}{36}
\affiliation{\TEMPLE}
\newcommand*{\JLAB}{Thomas Jefferson National Accelerator Facility, Newport News, Virginia 23606}
\newcommand*{\JLABindex}{37}
\affiliation{\JLAB}
\newcommand*{\UTFSM}{Universidad T\'{e}cnica Federico Santa Mar\'{i}a, Casilla 110-V Valpara\'{i}so, Chile}
\newcommand*{\UTFSMindex}{38}
\affiliation{\UTFSM}
\newcommand*{\EDINBURGH}{Edinburgh University, Edinburgh EH9 3JZ, United Kingdom}
\newcommand*{\EDINBURGHindex}{39}
\affiliation{\EDINBURGH}
\newcommand*{\GLASGOW}{University of Glasgow, Glasgow G12 8QQ, United Kingdom}
\newcommand*{\GLASGOWindex}{40}
\affiliation{\GLASGOW}
\newcommand*{\VT}{Virginia Tech, Blacksburg, Virginia   24061-0435}
\newcommand*{\VTindex}{41}
\affiliation{\VT}
\newcommand*{\VIRGINIA}{University of Virginia, Charlottesville, Virginia 22901}
\newcommand*{\VIRGINIAindex}{42}
\affiliation{\VIRGINIA}
\newcommand*{\WM}{College of William and Mary, Williamsburg, Virginia 23187-8795}
\newcommand*{\WMindex}{43}
\affiliation{\WM}
\newcommand*{\YEREVAN}{Yerevan Physics Institute, 375036 Yerevan, Armenia}
\newcommand*{\YEREVANindex}{44}
\affiliation{\YEREVAN}
\newcommand*{\NOWUK}{University of Kentucky, Lexington, Kentucky 40506}
\newcommand*{\NOWISU}{Idaho State University, Pocatello, Idaho 83209}
\newcommand*{\NOWSCAROLINA}{University of South Carolina, Columbia, South Carolina 29208}
\newcommand*{\NOWINFNGE}{INFN, Sezione di Genova, 16146 Genova, Italy}

\author {N. ~Hirlinger Saylor} 
\affiliation{\RPI }
\affiliation{\ORSAY }
\author {B. ~Guegan} 
\affiliation{\ORSAY }
\affiliation{\RPI}
\author {V.D.~Burkert} 
\affiliation{\JLAB}
\author {L.~Elouadrhiri} 
\affiliation{\JLAB}
\author {M.~Gar\c con} 
\affiliation{\SACLAY}
\author {F.X.~Girod} 
\affiliation{\JLAB}
\affiliation{\SACLAY}
\author {M.~Guidal}
\affiliation{\ORSAY }
\author {H.S.~Jo}
\affiliation{\KNU }
\affiliation{\ORSAY }
\author {V.~Kubarovsky} 
\affiliation{\JLAB}
\author {S.~Niccolai}
\affiliation{\ORSAY }
% ...

\begin{abstract} 
This paper reports the measurement of polarized and unpolarized cross sections for the $ep\rightarrow e'p'\gamma$ reaction, which is comprised of Deeply Virtual Compton Scattering (DVCS) and Bethe-Heitler (BH) processes, at an electron beam energy of 5.88 GeV at the Thomas Jefferson National Accelerator Facility using the  Large Acceptance Spectrometer CLAS. The unpolarized cross sections and polarized cross section differences have been measured over broad kinematics, $0.10 < x_{B} < 0.58$, $1.0 < Q^{2} < 4.8 \text{ GeV}^{2}$, and $0.09 < -t < 2.00 \text{ GeV}^{2}$. The results are found to be consistent with previous CLAS data, and these new data are discussed in the framework of the generalized parton distribution  approach. Calculations with two widely used phenomenological models, denoted VGG and KMSC, are approximately compatible with the experimental results over a large portion of the kinematic range of the data. \end{abstract}


\maketitle

\vskip 0.5cm

\section{Introduction}
While nucleons have been known to consist of quarks and gluons for nearly half a century, a quantitative three-dimensional description of that structure is still a topic of exploration and great interest. How the properties of quarks and gluons contribute to the nucleon is a fundamental question, whether it concerns their mass or spin, or their distributions in momentum or position space.

The traditional method for exploring nucleon structure has been through electron scattering off nucleons. Two complementary approaches have historically been taken, namely elastic scattering $eN\rightarrow e'N'$  \cite{Hofstadter56}, and inclusive scattering $eN\rightarrow e'X$ \cite{Breidenbach}, on the nucleon. Through elastic electron-nucleon scattering, form factors (FFs) may be obtained, whose Fourier transforms give access to the transverse spatial distributions with respect to the incident virtual photon direction. Through inelastic electron-nucleon scattering, parton distribution functions (PDFs) may be obtained, which describe the longitudinal momentum distributions.

While these methods have been effective in answering many questions regarding nucleon structure, a complete three-dimensional description is much desired. Two complementary approaches have been developed in the last few decades, namely through transverse momentum dependent parton distributions (TMDs), and generalized parton distributions (GPDs). 
See Refs.~\cite{Ji:1998pc,Belitsky:2005qn,Mueller:1998fv,Diehl:2003nv,Goeke:2001tz} for reviews. 

TMDs may be accessed through inclusive and semi-inclusive deep inelastic scattering (DIS), giving a description of the nucleon in terms  of longitudinal and transverse momentum. GPDs may be accessed via exclusive reactions such as deeply virtual Compton scattering (DVCS) $eN\rightarrow e'N'\gamma$ and exclusive meson production 
$ep\rightarrow e'N'M$. These processes describe the  nucleon in terms of the longitudinal momentum and transverse position of its quarks and gluons, which has often been described as nucleon tomography.

Beyond the 3D imaging, proton GPDs are related, through sum rules ~\cite{Ji:1996ek}, to the gravitational form factors of the proton, which themselves are related to mechanical properties of the proton~\cite{ Polyakov:2002yz}. These relations have recently been used to determine the pressure distribution inside the proton from experimental data ~\cite{Burkert:2018bqq}.

In this article, we focus on the study of DVCS, which provides the cleanest extraction of GPDs.
For the DVCS reaction, in the Bjorken limit, GPDs are extracted by means of  factorization of the DVCS amplitude into hard and soft parts. The hard part consists of the virtual Compton process at the quark level $\gamma^{*} q \rightarrow \gamma q$, which is  perturbatively calculable in QED.  The soft part is  nonperturbative, which, in  leading twist and leading order, is  parametrized in terms of four GPDs. In this framework, the amplitude is dominated by scattering on a single quark.   Detailed reviews on applications of GPDs to DVCS based on factorization are found in Refs.~ \cite{Goeke:2001tz,Diehl:2003nv,Belitsky:2005qn,Guidal:2013rya}.

\begin{figure}
\includegraphics[width=0.4\textwidth]{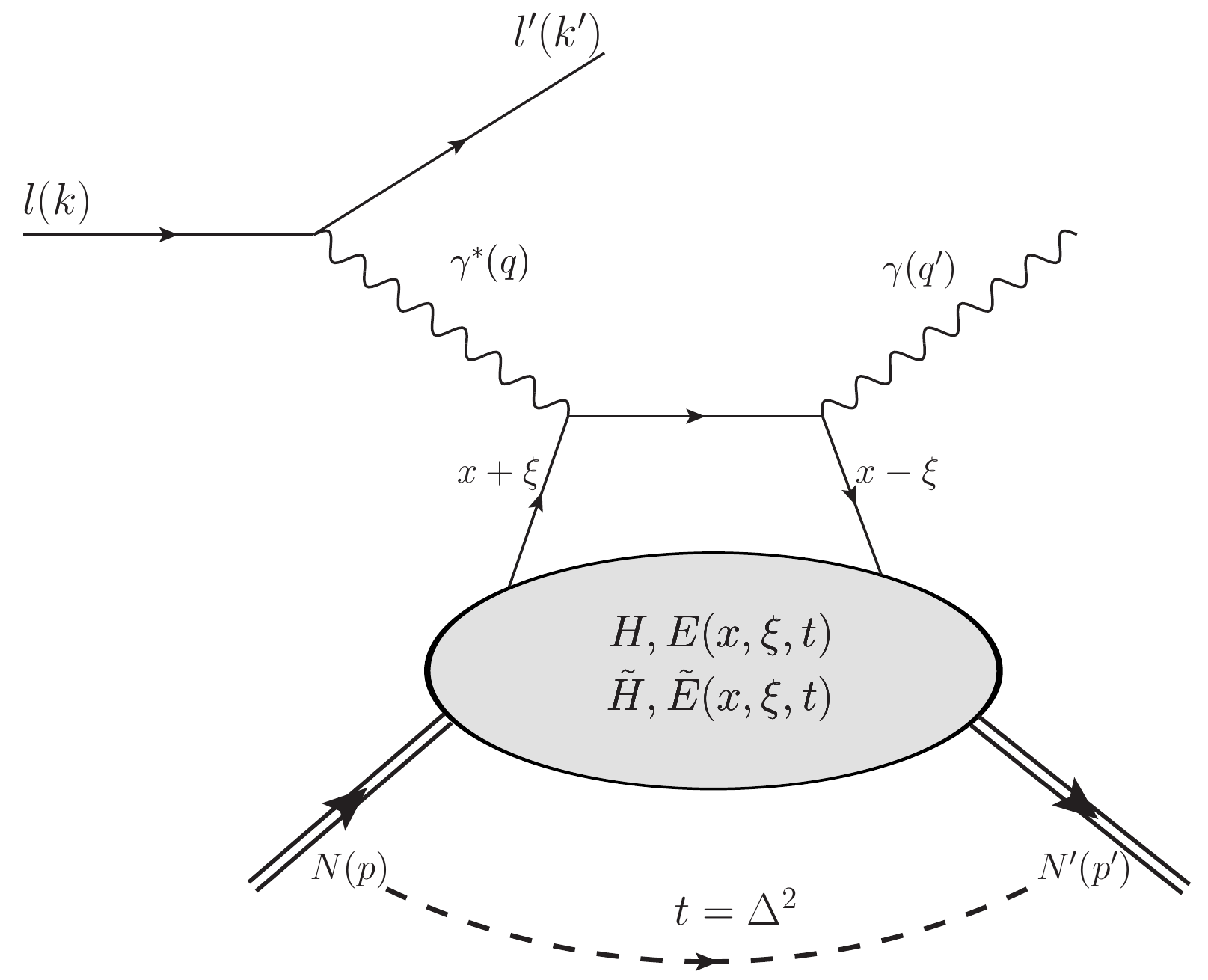}
\caption{The ``handbag'' diagram for the DVCS process on a nucleon. \label{fig:dvcs}}
\end{figure}

GPDs depend upon three variables: $x$, $\xi$ and $t$ (see Fig.\ref{fig:dvcs}). They represent the probability amplitude of finding a quark (anti-quark) in the nucleon with a longitudinal momentum fraction $x+\xi$ and of putting it back into the nucleon with a longitudinal momentum fraction $x-\xi$, plus a transverse momentum kick denoted by $\Delta$. $\xi$ is referred to as the skewness, and $-2\xi$ is the longitudinal momentum fraction of the transfer $\Delta$. In the Bjorken limit, $\xi$ is related to the standard deep inelastic Bjorken variable as follows: $\xi \thicksim \frac{x_{B}}{2-x_{B}}$, in which $x_{B} = \frac{Q^2}{2m_p(E-E^\prime)}$.    The quantity $m_p$ is the proton mass, $E$ is the beam energy and $E^\prime$ is the scattered electron energy.  The squared electron four-momentum transfer, is related to  the incoming and scattered electron four momenta as $Q^2 = (p_e -p_e^\prime)^2$. The variable $t=\Delta^{2}=(p_N'-p_N)^{2}$ is the squared four momentum transfer between the initial and final nucleons.

The lower blob of Fig.~\ref{fig:dvcs} can be described in terms of the four GPDs, $H$, $E$, $\tilde{H}$, $\tilde{E}$, which are called chiral-even GPDs because they conserve the helicity of the struck quark. The GPDs $H$ and $E$ are independent of the quark helicity and are therefore called unpolarized GPDs, whereas $\tilde{H}$ and $\tilde{E}$ depend on the quark helicity and are called polarized GPDs. $H$ and $\tilde{H}$ conserve the proton helicity, whereas $E$ and $\tilde{E}$  flip the proton helicity.

The GPDs obey some sum rules and have limits that are model-independent. The first moment in $x$ of 
$H$, $E$, $\tilde{H}$ and $\tilde{E}$ give the $F_{1}(t)$, $F_{2}(t)$, $G_{A}(t)$ and $G_{P}(t)$ form factors, 
which are the Dirac, Pauli, axial and pseudoscalar form factors, respectively~\cite{Ji:1996nm}. 
In the forward limit,  $\xi$  and $t\to 0$, the $H$ and $\tilde H$ GPDs are related  to the unpolarized and polarized PDFs, respectively.
Also, Ji derived a sum rule~\cite{Ji:1996ek} that links the second moment in $x$ of the GPDs $H$ and $E$ in the forward limit to the total (spin + orbital) angular momentum carried by the quarks.

It is possible to interpret the GPDs as an $x$-decomposition of the form factors, where we have access to the FFs at different values of $x$, rather than integrated over $x$. As $\xi \to 0$, the 2-dimensional Fourier transform of the GPD $H_{q}(x,0, t)$ is linked to the distribution $q(x,b_{\perp})$, 
which provides a simultaneous measurement of the longitudinal momentum $x$ at a given transverse position (impact parameter $b_{\perp}$) for unpolarized 
quarks and target~\cite{Burkardt:2000za}: 

\begin{equation}
q(x,0,b_{\perp})=\int \frac{d^{2}\Delta_{\perp}}{(2\pi)^{2}}e^{-ib_{\perp}\Delta_{\perp}}H^{q}(x,0,-\Delta_{\perp}^{2}).
\end{equation}

The information contained in the PDFs of the deep inelastic scattering (DIS) experiments and the form factors  measured in elastic scattering are now unified into the GPD framework, allowing for a three dimensional (2 spatial and 1 momentum) picture of the nucleon.

The four GPDs ($H, \tilde{H}, E, \tilde{E}$) depend on the three variables $x$, $\xi$, and $t$, but only $\xi$ and $t$ are accessible experimentally. Thus, one cannot directly access the $x$-dependence of the GPDs from DVCS experiments. In  DVCS experiments one accesses the integral of the GPD over $x$. This integral is called a Compton Form Factor (CFF):

\begin{equation}
\mathcal{H}(\xi, t) = \int^{+1}_{-1}dx H(x,\xi,t)\left(\frac{1}{\xi - x - i \epsilon} - \frac{1}{\xi + x - i\epsilon}\right).
\label{eq:cff}
\end{equation}

The CFFs depend only on $\xi$ (or equivalently on $x_{B}$) and $t$, and are the quantities that can be extracted from DVCS experiments. Model-independent fitting procedures have been developed which, for a given experimental point ($x_{B}, t$), keep the CFFs as free parameters and extract them from the DVCS experimental observables. 

In addition to the DVCS amplitude, the cross sections for the exclusive electroproduction of a photon $eN\rightarrow e'N'\gamma$ also receive a contribution from the Bethe-Heitler (BH)  process, which is exactly theoretically calculable. In this case, the real photon is emitted either by the incoming or by the scattered electron but not by the nucleon itself, unlike for DVCS. This BH process does not give access to GPDs, but is experimentally indistinguishable from the DVCS process and interferes with it.

Different observables from the DVCS process must be measured to extract the information related to the GPDs. For instance, a polarized beam allows one to measure two independent observables: the unpolarized cross sections $\sigma_{unp}={1\over 2}(\sigma^{\rightarrow}+\sigma^{\leftarrow}$) and the difference of polarized cross sections $\Delta\sigma_{pol}={1\over 2}(\sigma^{\rightarrow}-\sigma^{\leftarrow})$ for opposite beam helicities.

The unpolarized and polarized cross sections, in terms of the DVCS and Bethe-Heitler amplitudes,  $\mathcal{M}_{DVCS}$ and  $ \mathcal{M}_{BH}$, are:
\begin{eqnarray}
	\sigma_{unp} 	&\propto& 	|\mathcal{M}_{BH}|^{2} \, +Re(\mathcal{M}_{int})+ \, |\mathcal{M}_{DVCS}|^{2} \nonumber \\
	\Delta\sigma_{pol}  	&\propto&  2 \, Im(\mathcal{M}_{int})\,
\label{Observables}
\end{eqnarray}
\noindent where the interference amplitude is
 $$\mathcal{M}_{int}=\mathcal{M}_{BH}\cdot \mathcal{M}^*_{DVCS}+\mathcal{M}^*_{BH}\cdot \mathcal{M}_{DVCS}$$
The real(imaginary) part of the interference amplitude is proportional to the real(imaginary) part of the Compton form factor.

\begin{figure}
\includegraphics[width=0.5\textwidth]{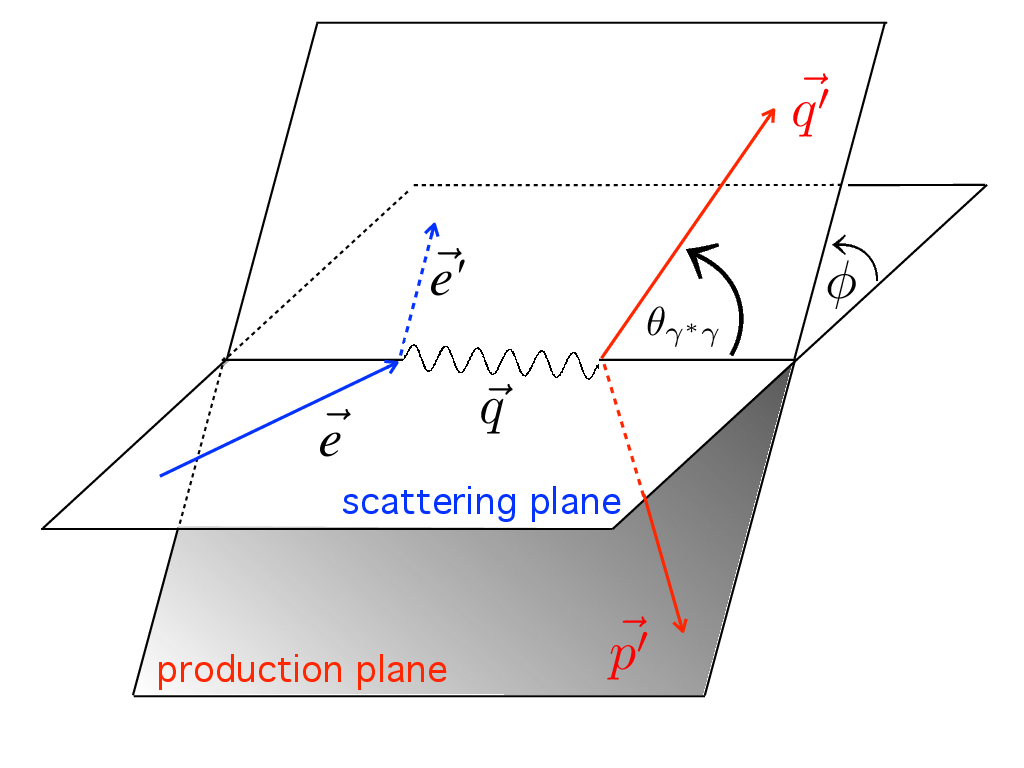}
\caption{The electron scattering plane and the plane of the produced photon ${\vec q}\prime$, and recoil proton, ${\vec p}\prime$, respectively, in DVCS. The angle between the photon plane and electron plane is  $\phi$. (Color online.)}
\label{fig:had}
\end{figure}

The term $|{\mathcal{M}}_{DVCS}|^{2}$ is significantly smaller than  $|{\mathcal{M}}_{BH}|^{2}$, 
except near  $\phi = 180^\circ$, where ${|\mathcal{M}}_{DVCS}|$ can be larger than ${\mathcal{M}}_{BH}$. Here, $\phi$ is the photon azimuthal angle relative to the electron scattering plane, as in Fig.~\ref{fig:had}.  
Thus, the observable $\sigma_{unp}$  gives  access to the the combination $Re( \mathcal{M}_{DVCS})+ \, |\mathcal{M}_{DVCS}|^{2}$ primarily  near $\phi$ = 180$^\circ$. 
Since the BH process off of an unpolarized nucleon does not exhibit a beam polarization asymmetry, the observable $\Delta\sigma_{pol}$  gives access to the imaginary part of the DVCS amplitude,  $Im(\mathcal{M}_{DVCS})$, which in the twist-2 approximation is a combination of the GPDs at the point $x=\pm \xi$.

Several programs around the world,  at Hall A
~\cite{Defurne:2015kxq,MunozCamacho:2006hx} and CLAS
~\cite{Stepanyan2001,Chen:2006na,FXGirod2008,Seder:2014cdc,Pisano:2015iqa,Jo:2015ema} at Jefferson Laboratory, COMPASS~\cite{GautheronCOMPASS,COMPASS},   and HERA~\cite{SChekanov2009,AAktas2005,FDAaron2008,CAdloffetal2001}, have been seeking to explore GPDs through the use of DVCS and other channels such as exclusive meson leptoproduction.
We focus here on DVCS on the proton, and the polarized and unpolarized cross section observables, as
measured with the 6 GeV electron beam of Jefferson Laboratory.

\section{Experiment}

The measurement of the $ep\rightarrow e'p'\gamma$ cross sections was conducted using the Continuous Electron Beam Accelerator Facility (CEBAF) Large Acceptance Spectrometer (CLAS)~\cite{CLASref} in Hall B of the Thomas Jefferson National Accelerator Facility (Jefferson Laboratory) in Newport News, VA, which was capable of delivering up to 6 GeV electrons to three experimental halls. The CLAS detector was based on a six-sector toroidal magnet, whose six coils were placed symmetrically around the beam line.

\begin{figure}
\includegraphics[width=0.5\textwidth]{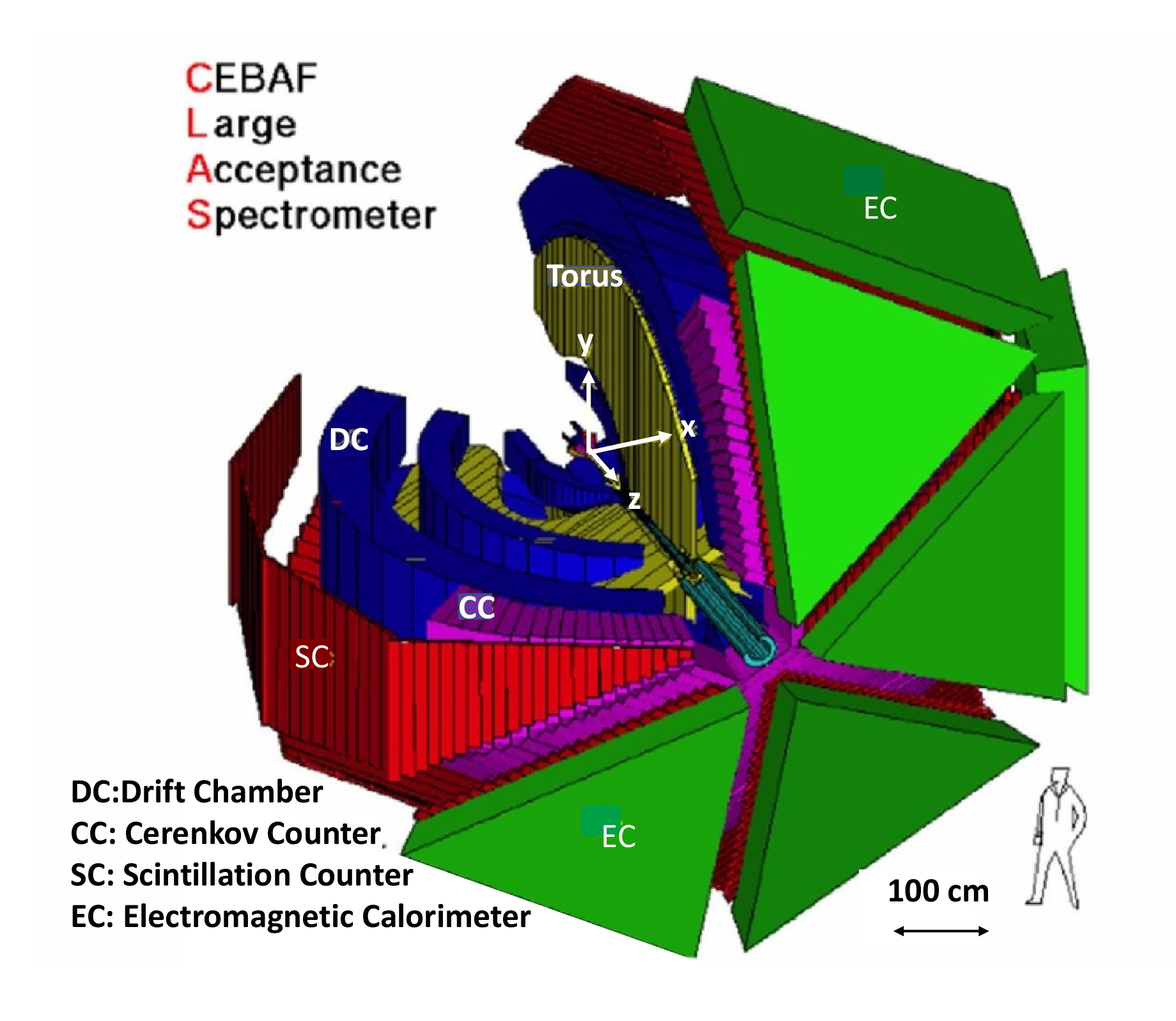}
\caption{A diagram of the CLAS detector, showing each of the six detectors. (Color online.)}
\label{fig:clas}
\end{figure}

Identification, timing, and tracking of charged particles was accomplished by three regions of drift chambers (DC) and by scintillator counters (SC) using the time-of-flight method. \v{C}erenkov counters (CC) and electromagnetic calorimeters (EC) were responsible for the identification of electrons. In addition, the EC was also responsible for the identification of photons and neutrons. Each of these sub-detectors is displayed in Fig.~\ref{fig:clas}.

\begin{figure}
\includegraphics[width=0.5\textwidth]{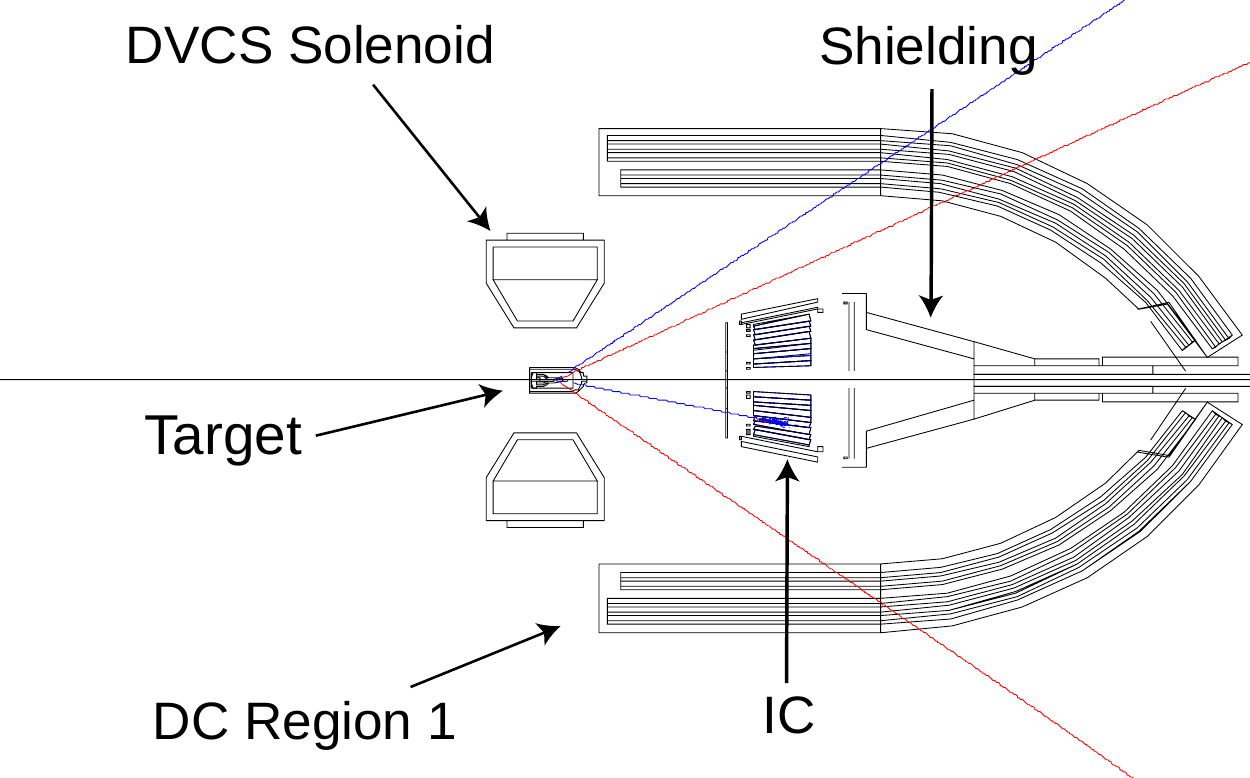}
\caption{Magnification of the CLAS target region within the innermost drift chamber, DC-region 1. Also shown is the inner calorimeter IC . }
\label{fig:IC}
\end{figure}

The data used in this analysis were obtained from an experiment called e1-dvcs2, which ran from  October 2008  to  January 2009, for a total of 90 days of run time, and a total integrated luminosity of $4.5 \times 10^7$ nb$^{-1}$. An 85\% polarized electron beam of energy 5.88 GeV was incident on a 5 cm liquid hydrogen target, held at 20 K, and placed at 57.5 cm upstream relative to the center of the CLAS detector. In addition to the standard CLAS detector package, e1-dvcs2 ran with a forward angle inner calorimeter (IC)(see Fig.~\ref{fig:IC}), which was composed of PbWO${}_{4}$ crystals, and whose front face was placed at the center of CLAS. The IC enabled the detection of DVCS photons in the forward direction, where they are more common. The experiment was also equipped with a solenoid, in which the target was placed. The   solenoid  swept the M{\"o}ller background created during electron beam interactions in the target into the beam pipe.

\begin{figure}
\centering
\includegraphics[width=0.47\textwidth]{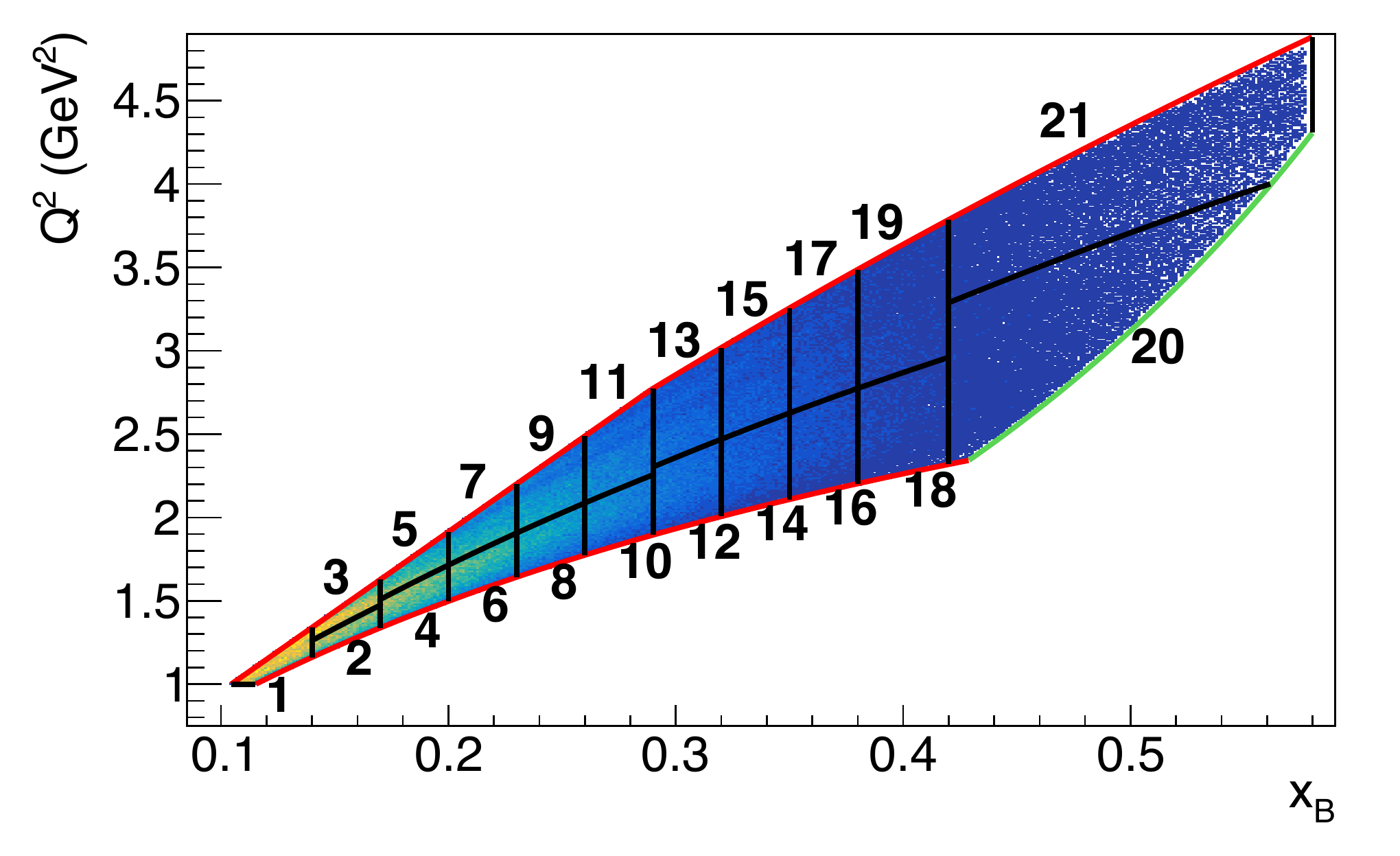}
\caption{The $Q^2$ vs. $x_B$ kinematical domain for e1-dvcs2 after all cuts. The heavy black lines correspond to the bin definitions.}
\label{fig:KinematicalDomain}
\end{figure}

Here, we report the measurement of the unpolarized DVCS differential cross sections $\sigma_{unp}$ and the beam polarization cross section 
differences $\sigma_{pol}$ as a function of $Q^2,  x_B, t$, and $\phi$.
Kinematically, the 3-body final state reaction $ep\rightarrow e'p'\gamma$ depends on these four independent variables.
The kinematic coverage for this reaction is shown in Fig.~\ref{fig:KinematicalDomain}. For the purpose of physics analysis an additional cut  removed events with $W > 2$~GeV, where $W$ is the $\gamma^*p$ center-of-mass energy.

The differential cross section is related to the measured quantities according to:

\begin{equation}
\frac{d^4 \sigma_{ep \rightarrow e^\prime p^\prime \gamma}}{dQ^2 dx_B dt d\phi} = 
 \frac{N(Q^2,x_B,t,\phi)}{ \Delta Q^2  \Delta x_B \Delta t \Delta \phi} 
\frac{1}{\mathcal{L}_{int}\epsilon_{ACC}\delta_{RC} \delta_{Norm}}.
\label{eq:sig_ep_eppippim}
\end{equation}

\noindent The definitions of the  quantities in Eq.~\ref{eq:sig_ep_eppippim}  are:

\begin{itemize}

\item $N(Q^2,x_B,t,\phi)$ is the  number of $ep \rightarrow e'p'\gamma$ events in a given $Q^2,x_B,t,\phi$ bin;

\item $\mathcal{L}_{int}$ is the integrated luminosity (which takes
into account the correction for the data-acquisition dead time);

\item $(\Delta Q^2 \Delta x_B \Delta t \Delta \phi)$ is the corresponding bin width, including a 4-dimensional correction to take into account partial occupation of the phase space originating from various cuts.
The specified $Q^2$, $x_B$, and $t$  values are the mean values of the data for each variable for each 4-dimensional bin.

\item $\epsilon_{ACC}$ is the acceptance calculated for each bin $(Q^2,x_B,t,\phi)$;

\item $\delta_{RC}$ is the correction factor due to the radiative effects calculated for each $(Q^2,x_B,t,\phi)$ bin;

\item $\delta_{Norm}$ is an overall absolute renormalization factor calculated from the elastic cross section measured in the same experiment (see Sec.~\ref{systematics});
 
\end{itemize}

The kinematic binning of the experimental data, shown in Fig.~\ref{fig:KinematicalDomain}, was determined by the physics requirement, which is the study of the reaction in the deep inelastic region -  $W>2\ \text{GeV}$ and $Q^{2} > 1\ \text{GeV}^{2}$, as constrained by experimental acceptances and capabilities, as well as other  physical kinematic constraints:

\begin{itemize}
\item The polar angular acceptance of electrons was from $21^{\circ}$ to $45^{\circ}$.
\item The scattered electron minimum energy was $0.8 \ \text{GeV}$.
\item $-t  > -t_{min}$, where $-t_{min}$ is the minimum physical $-t$  for a given $x_B$ and $Q^2$.

\item We chose to place a cut at $\theta_\gamma > 4.77$ deg. for two reasons. 

1. To avoid any low angle area not covered by the IC.

2.  Additionally, since the BH cross section is nearly singular at $\theta_{\gamma} \rightarrow 0$ deg., This cut allowed us to avoid relying only on the event generator whose cross sections sections varied rapidly in this region.

\end{itemize}

There were 21 bins in $Q^2$ and $x_B$, 9 bins in $-t$, and 24 bins in $\phi$. Therefore, there are a total of 189 angular distributions as a function of $\phi$. The binning is presented in table form in the Appendix.
\section{Particle Identification}

To ensure the exclusivity of the $ep\rightarrow e'p'\gamma$ process and minimize background, we identified the three
particles of the final state of the reaction, i.e. we accepted every event which has one electron, one proton, and at least one photon.

\subsection{Electron Identification}

The identification of electrons required  the detection of  a negatively charged particle in the CC and EC, in the same sector of CLAS, with a momentum greater than 0.8 GeV  This trigger  suppressed pions with momentum up to 2.5 GeV, with an effective suppression of 100\% for momenta less than 0.8 GeV.  Since electrons lose nearly all of their energy through showering, they deposit an energy in the EC proportional to their momentum. The sampling fraction is defined as $f_s = E_{\text{total}}/p$, where $E_{total}$ is the total energy deposited in the EC, and $p$ is the momentum as measured with the DCs by the curvature of the particle's trajectory in the CLAS toroidal field. 

\begin{figure}[!ht]
\includegraphics[width=0.47\textwidth]{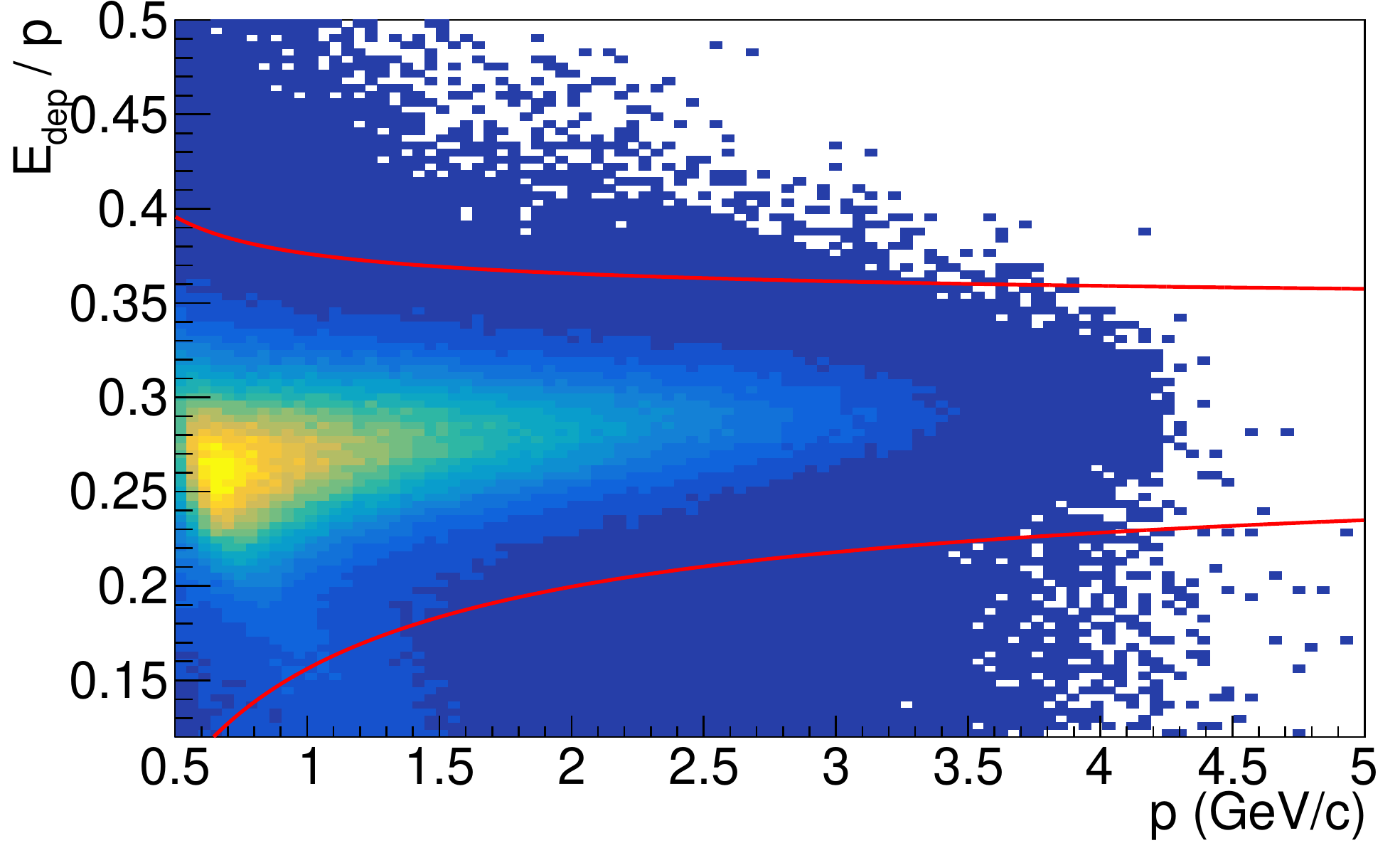}
\caption{Sampling fraction $f_s$ as a function of momentum, with the electron selection cut shown as a solid red curve.  The momentum dependence is visible from this plot. An artefact of data processing is visible at  $f_s = 0.12$, which lies  outside of the electron selection cut. (Color online.)}
\label{fig:p_samp}
\end{figure}

More massive negatively charged pions lose a greater  fraction of their energy via ionization, which is  a smaller fraction of their energy than electrons lose in the EC, so that their sampling fraction was distributed significantly less than for electrons. The sampling fraction vs. momentum is shown in Fig.~\ref{fig:p_samp}.  The resolution was momentum dependent, broadening at lower momentum. The final cut, indicated by  solid red curves, required that a particle had a sampling fraction that lay within three standard deviations of the mean  $f_s$,  which is a function of momentum. Additionally, fiducial cuts were applied to avoid areas where the geometrical acceptance of CLAS was small and changing rapidly.

\subsection{Proton Identification}

Protons are identified by comparing the velocities $\beta = v / c$ of the proton candidates (all positive particles) as measured from the DC and TOF:

\begin{equation}
\Delta \beta = \beta_{TOF} - \beta_{DC}.
\label{eq:tof}
\end{equation}

\begin{figure}[!ht]
\includegraphics[width=0.47\textwidth]{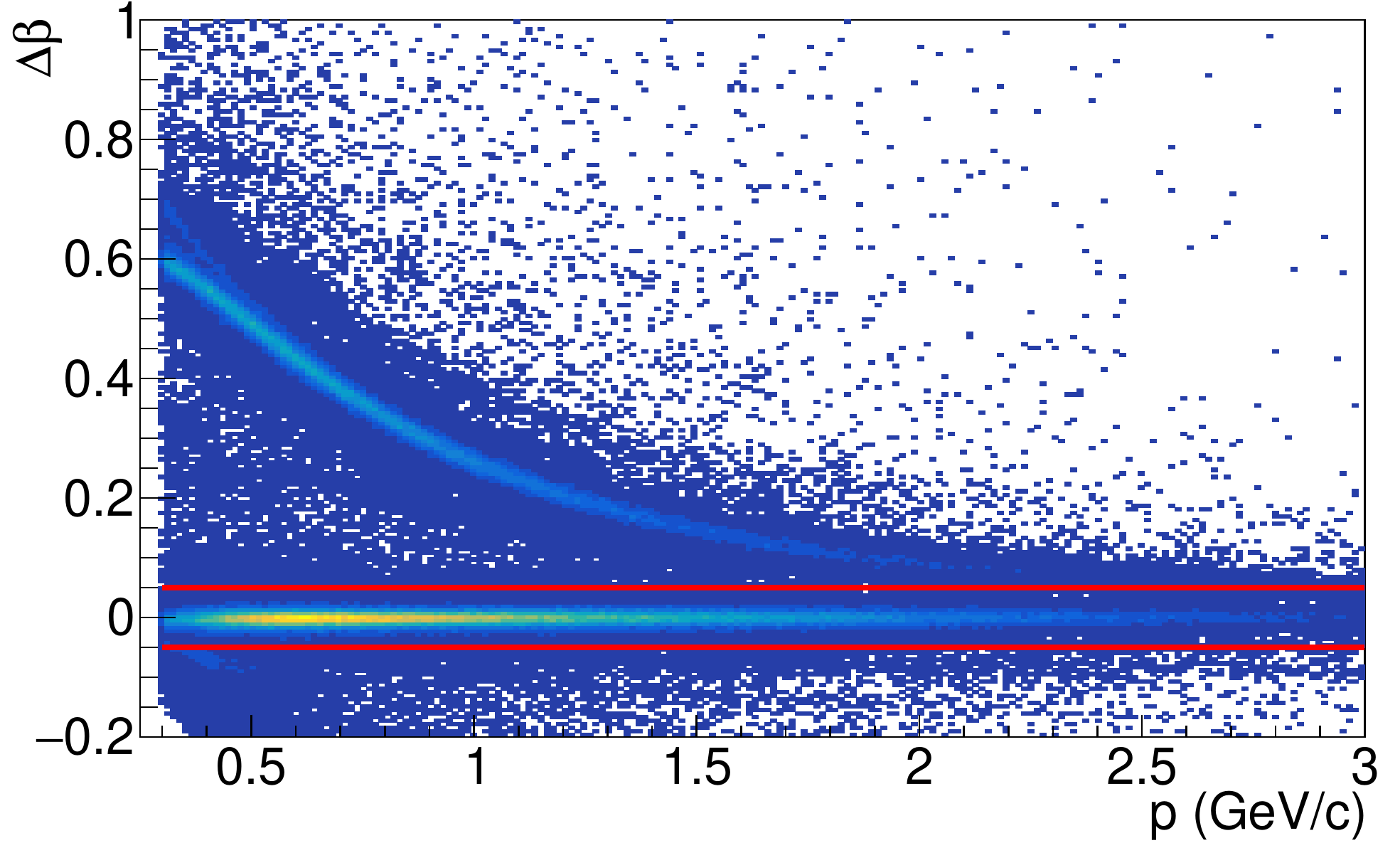}
\caption{The proton identification cut $\Delta\beta$ as a function of momentum $p$. The  solid horizontal red lines  represent the restriction that $|\Delta\beta| < 0.05$. This serves to separate the protons from other particles, which are visible in this plot. (Color online.)} 
\label{fig:deltab_p}
\end{figure}

Since many variables required a momentum dependent cut, we plot this variable as a function of $p$. A visual inspection is enough to determine that the width of the peak about $\beta = 0$ does not significantly  broaden at any region. Therefore a horizontal cut  around $\Delta\beta =0$ was taken. The cut chosen for this analysis was $|\Delta\beta| < 0.05$. A plot illustrating this cut is shown in Fig.~\ref{fig:deltab_p}.
As with the electrons, the detectors involved with proton identification included geometrical fiducial cuts to reject regions of poorly known detector behavior.

\subsection{Photon Identification}

Two detectors in this experiment were used  to identify photons. The first was the EC for lab scattering angles  $\theta > 20^{\circ}$ and the second was the IC, for $\theta < 20^{\circ}$.

The identification of photons in the EC was achieved by making use of EC timing information, in addition to an absence of any associated track  in the DC. The reconstructed relativistic velocity $\beta_{rec}$ of the photon candidate is defined as:
    
\begin{equation}
\beta_{\text{rec}} = \frac{|\vec{r}|}{c(t_{\text{EC}} - t_{\text{tr}})} \text{,\ \ with\ \ } \vec{r} = \vec{e} - \vec{v},
\label{eq:betarec}
\end{equation}

\noindent where $t_{\text{EC}}$ is the timing in the EC relative to the reference time (trigger time) $t_{\text{tr}}$. The vector $\vec{e}$ extends from the  center of CLAS to the hit in the EC and $\vec{v}$ is the vector running from the  CLAS center to the corrected vertex position of the electron that was detected in coincidence.

\begin{figure}[!ht]
\centering
\includegraphics[width=0.45\textwidth]{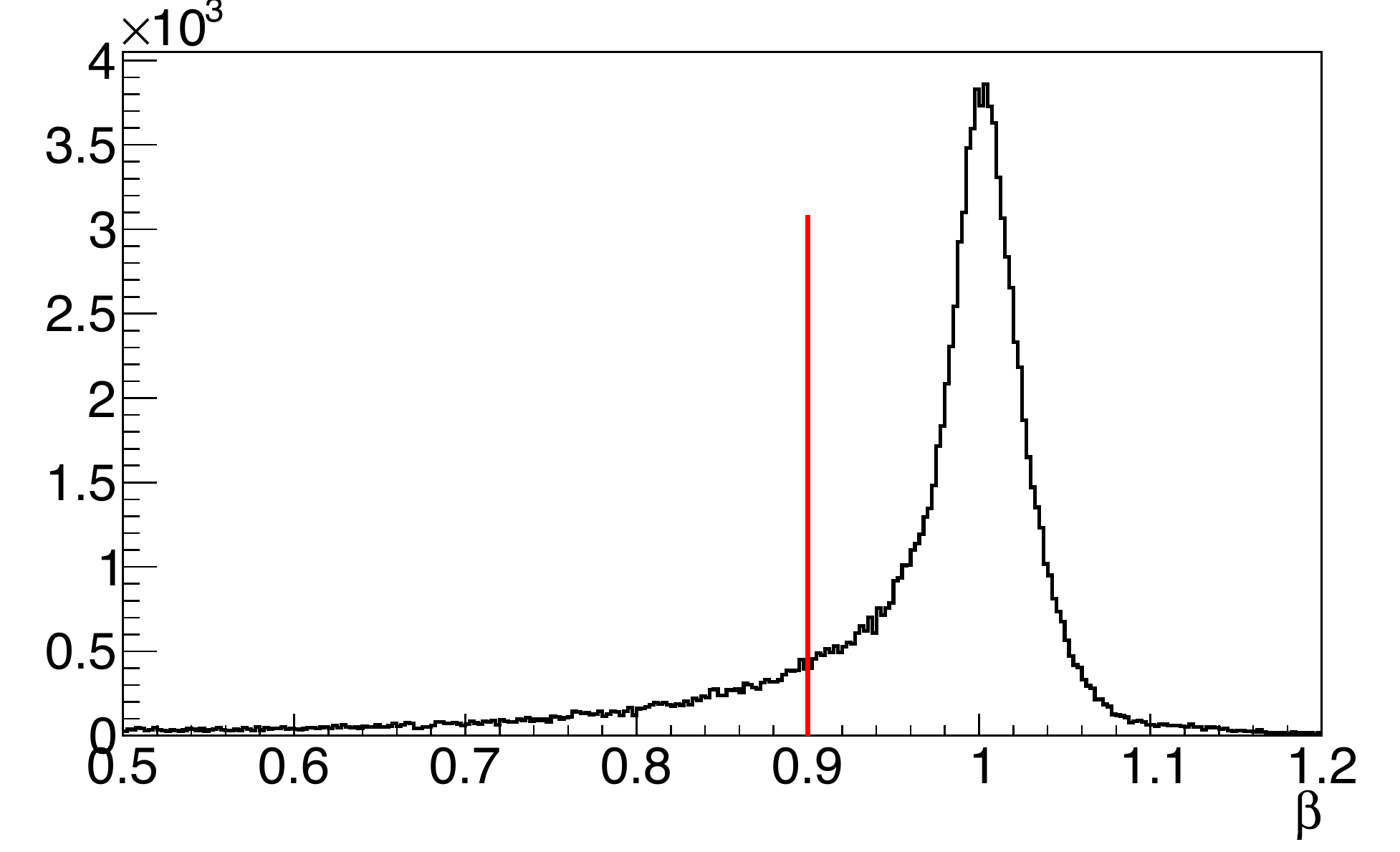}
\caption{$\beta_{rec}$ for photon candidates, with a cut accepting only particles with $\beta > 0.9$.}
\label{fig:b_photon}
\end{figure}

 As expected, one sees a peak at $\beta = 1$, representing a distribution of photons, and a tail   towards  lower $\beta$, representing the neutrons. The final photon selection cut requires $\beta > 0.9$ (see Fig.~\ref{fig:b_photon}).

Most of the background in the IC originated from M{\o}ller electrons. These were efficiently removed in the analysis through a correlated small-angle/low-energy cut. After this cut was applied, it was assumed (and checked a posteriori with the exclusivity cuts discussed below) that any signal in the IC was from a photon.

\section{Corrections to Data}

Several corrections to the measured experimental variables were applied.  
In particular, the beam energy was determined from elastic scattering events and, for the $ep\gamma$ events, the electron and proton momenta and angles were corrected. Details regarding these corrections are provided in the following subsections.
\subsection{Energy Loss Correction}

 Corrections for electron and proton energy loss occurring in the target due to Bremsstahlung and ionization were carried out using the code ``GSIM'' (GEANT SIMulation),  to model these effects in CLAS. GSIM is based on the GEANT3 library developed at CERN~\cite{GEANT}.   This code simulated the response of the CLAS detector
 to the interactions of traversing particles, as  described in Sec.~\ref{sec:mc}.
 Events were generated at various positions within the target, with momentum $p_{\text{gen}}$ and angles  $\theta_{\text{gen}}$  and $\phi_{\text{gen}}$.
The events were then tracked through the detector, taking into account  the detector response, to yield the reconstructed momenta  and angles $p_{\text{rec}}$, $\theta_{\text{rec}}$, and $\phi_{\text{rec}}$.
The energy loss corrections were given by the difference between these two sets of momentum values, 
$$
\Delta p_{e} = p_{\text{gen},e} - p_{\text{rec},e},\text{\ and\ }
\Delta p_{p} = p_{\text{gen},p} - p_{\text{rec},p}.
$$

Fits to $\Delta p_{e}$  and  $\Delta p_{p}$ were carried out as  parametric functions of $\theta_e$ and  $\theta_p$ respectively. The corrections were then applied to the reconstructed  momenta in data.

\subsection{Kinematical Corrections}

Other corrections, beyond energy loss,  due to beam or detector misalignments or imperfect knowledge of the magnetic fields, were taken into account.  
We adopted a similar strategy as above, analyzing the measured W distributions and the reconstructed beam energy distribution from data, after the application of the energy loss corrections, and comparing them to the expected values of $W = m_p$ for beam energy $E_0 = 5.88$ GeV for the elastic channel.  For elastic scattering, the beam energy is related to the  electron and proton polar angles $\theta_{e}$ and $\theta_{p}$ as  

$$E_{0} = \frac{m_{p}}{\tan({\frac{\theta_{e}}{2}})\tan{(\theta_{p,})}} - m_{p}.$$

After the application of the energy loss corrections, the  $W$  and  beam energy distributions were reconstructed  for the elastic channel data using the empirical measurements  of $p_e$, $\theta_e$, $p_p$, and $\theta_p$,  and compared with the expected values of $W = m_{p}$  and  $E_{0} = 5.88 ~\text{ GeV}$.  From these differences, correction distributions  for electron momentum and angle as a function of $\theta_{\text{DC}}$ and $\phi_{\text{DC}}$, as measured in the first region of the DC, were obtained to better match the observed $W$ and $E_0$.

Because of the one-to-one correspondence of electron polar angle to proton polar angle in elastic scattering, the proton acceptance was limited to larger angles. This was due to the corresponding electrons at low angles being blocked by the IC. For those kinematics, corrections were made using  the 
$ep\rightarrow e'n\pi^{+}$ channel.  The corrections were made by comparing distributions of the ratios of pion momenta,  $p_{\pi^{+}}$, and energy, $E_{\pi^{+}}$,  with the measured values,  in terms of  the angles  measured in the drift chambers, $\theta_{\text{DC},\pi^{+}}$ and $\phi_{\text{DC},\pi^{+}}$.

These corrections very effectively brought the reconstructed values closer to the expected values. We conclude  that the corrections applied for both ionization corrections and kinematic corrections were done correctly based on the use of simulations and  of $E_{0}$, $W$ and $m_p$ as benchmarks. The corrections on average were less than 2\%.

\section{DVCS Analysis}

\label{sec:var}

After particle identification, a selection of the exclusive $ep\rightarrow e'p'\gamma$ channel was carried out. 
First, we demanded the presence of an electron, a proton, and at least one photon. Then we imposed constraints due to conservation laws - the so-called ``\textit{exclusivity}'' cuts. The following section outlines the list 
of variables used in such ``\textit{exclusivity}'' cuts, and the methods for determining the cuts on these variables.

These variables are:
\begin{enumerate}
\item $MM^{2}_{e'+p'}$: the squared missing mass  of the $ep\rightarrow e'p' X$ system,
\item $E_{X}$: the missing energy in the $ep\rightarrow e'p'\gamma X$ reaction,

\item $p_{T}$: the perpendicular component of missing momentum in the $ep\rightarrow e'p'\gamma X$ reaction,

\item $\theta_{\gamma,X}$: the difference between the calculated polar angle of the photon from the scattered electron and the
recoil proton measured from the kinematics of $ep\rightarrow e'p' X$ and the measured angle in $ep\rightarrow e'p'\gamma$, 

\item $ \phi_C$: the coplanarity of the virtual photon, the real photon, and the recoil proton. 
\end{enumerate}

Due to resolution effects, each variable had a distribution around the expected value. Monte Carlo and data distributions were analyzed in parallel. The method of choosing the exclusivity cuts was done in three stages:

\begin{enumerate}
\item Wide initial cuts were applied to the exclusivity distributions to clean up the data and Monte-Carlo distributions. The data and Monte-Carlo distributions  after these initial cuts are shown in Figs.~\ref{fig:zcut2017_0} and ~\ref{fig:zcut2017_1}. The upper panels and the  lower panels  are the data and Monte Carlo distributions, respectively. The data distributions exhibit strong peaks, as expected, superimposed on large backgrounds. The peaks in the data are also broader than those of the respective Monte Carlo distributions. 

\item The regions of the peaks in these  distributions were fit with empirical functions, representing the peaks, plus linear backgrounds under the peak regions; Gaussian for $MM^{2}_{e'+p'}$,  skewed Gaussian for  $E_{X}$,  and  custom functions  

$$ \theta_{\gamma,X} - A\sin{(x\sigma)}e^{-0.5(\frac{55\tan{(x\mu)}}{\sigma})^2}.$$

for $N(p_T)$ and $\theta_{\gamma,X}$.
The results of these fits are shown as curves superimposed on the data in Figs.~\ref{fig:zcut2017_0} and ~\ref{fig:zcut2017_1}. The vertical lines in these figures correspond to one and three standard deviations, i.e. $\sigma$ and  3$\sigma $, from the peak positions.

\item The final cuts, made at 3$\sigma$, were determined after inspection of the  results of  procedures 1 and 2 above.  The data and Monte-Carlo distributions after all cuts are presented in  Figs.~\ref{fig:zcut2017_2} and ~\ref{fig:zcut2017_3}.

Figure~\ref{fig.coplanarity} shows the coplanarity distribution, $\phi_C$,  for data and Monte-Carlo distributions after all  cuts on $p_{T}$, $E_{X}$, $MM^{2}_{e'+p'}$, $\theta_{\gamma,X}$.  It appears that the cuts on the other four variables are sufficient to obviate a cut on  $\phi_C$.

\end{enumerate}

\newpage
\begin{widetext}
\FloatBarrier
\begin{figure}[!ht]
\centering
\includegraphics[width=3.5in]{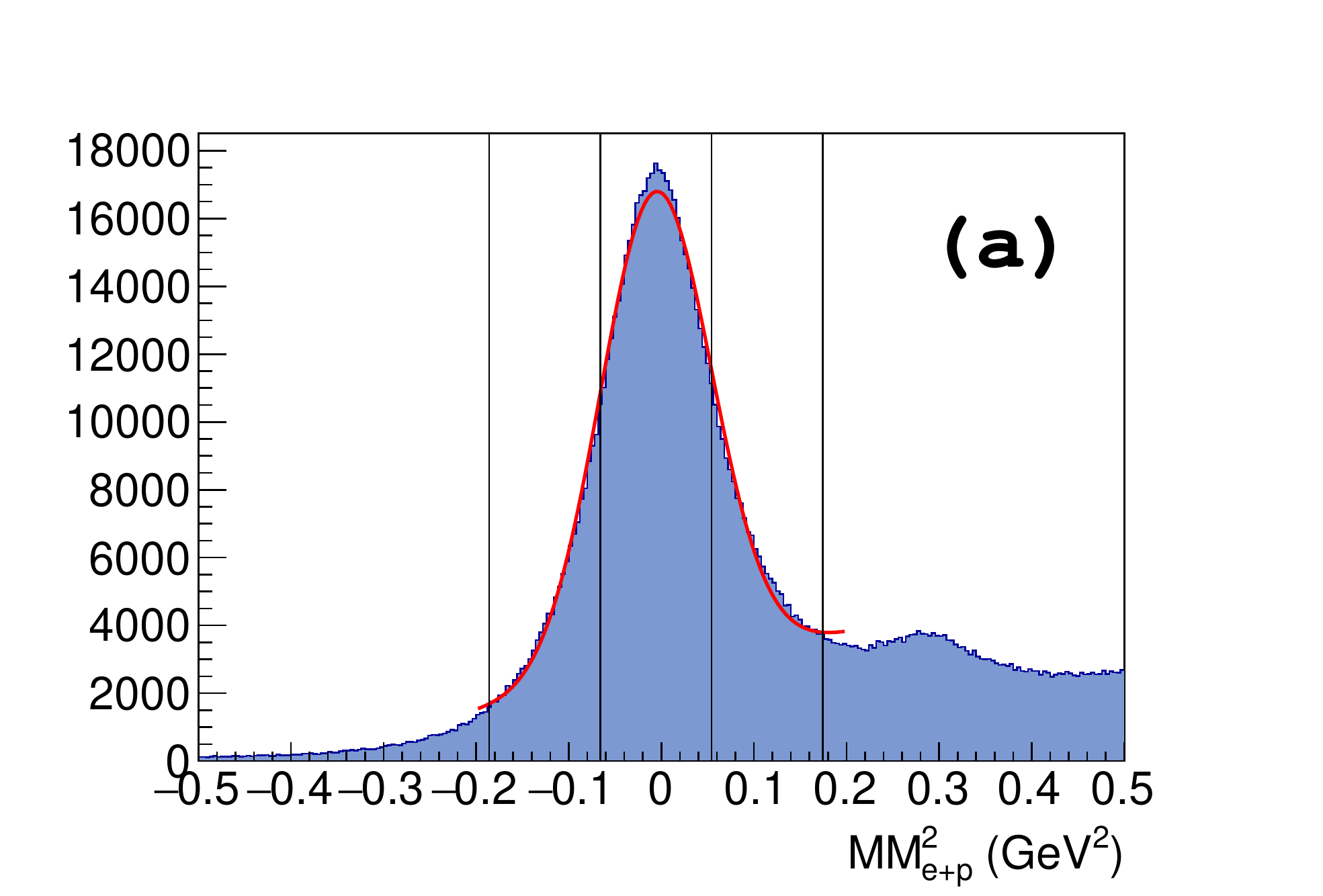}
\includegraphics[width=3.5in]{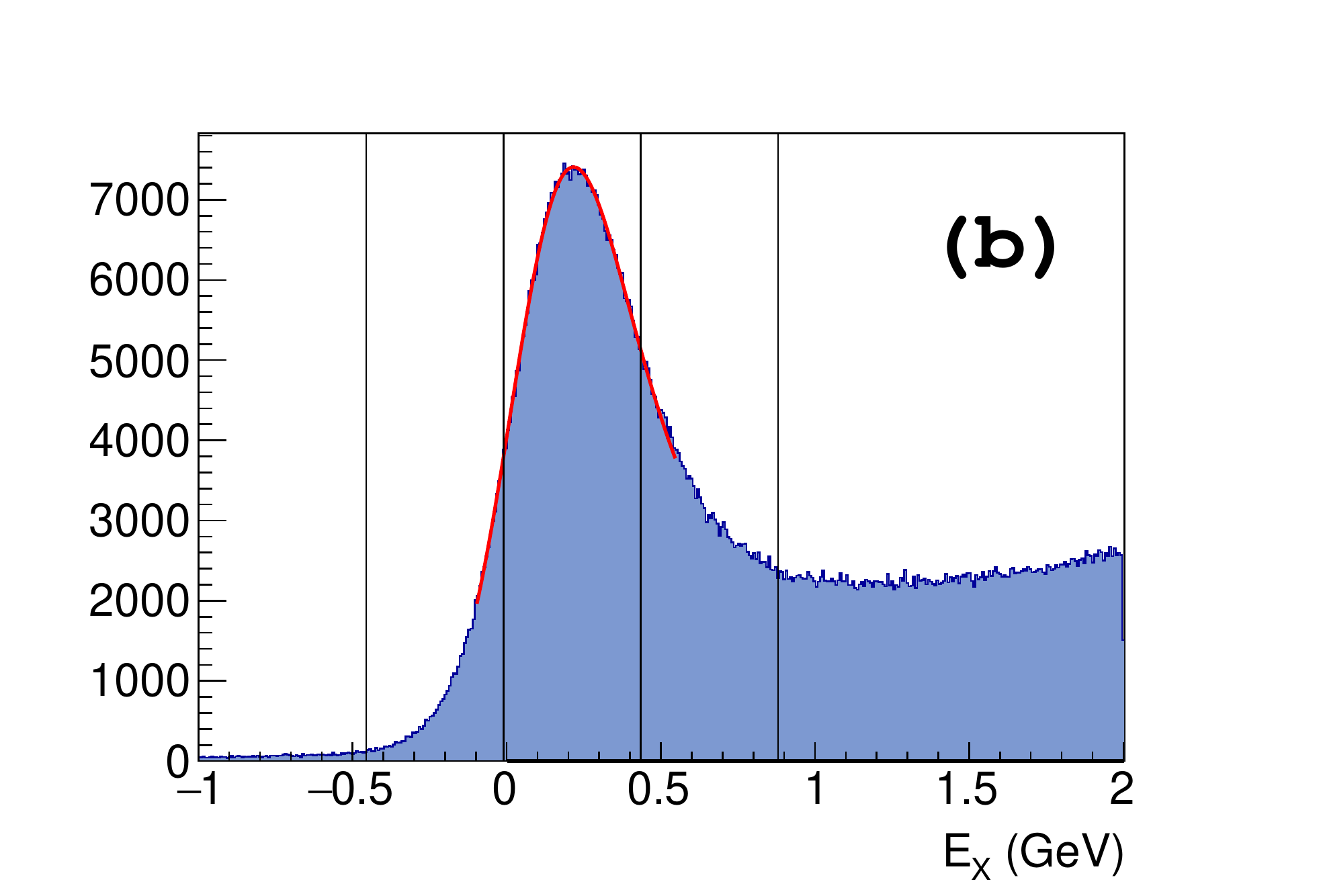}
\includegraphics[width=3.5in]{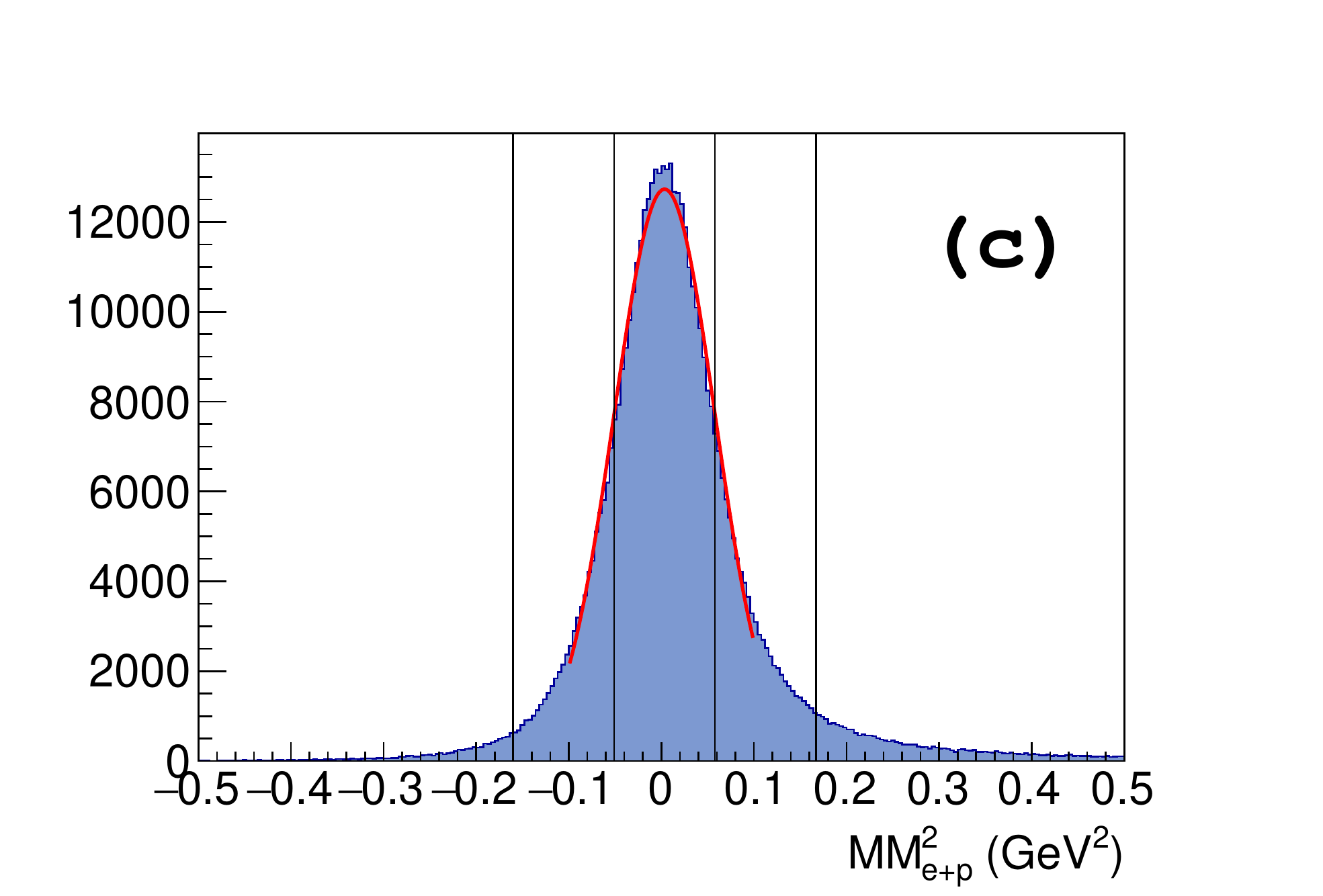}
\includegraphics[width=3.5in]{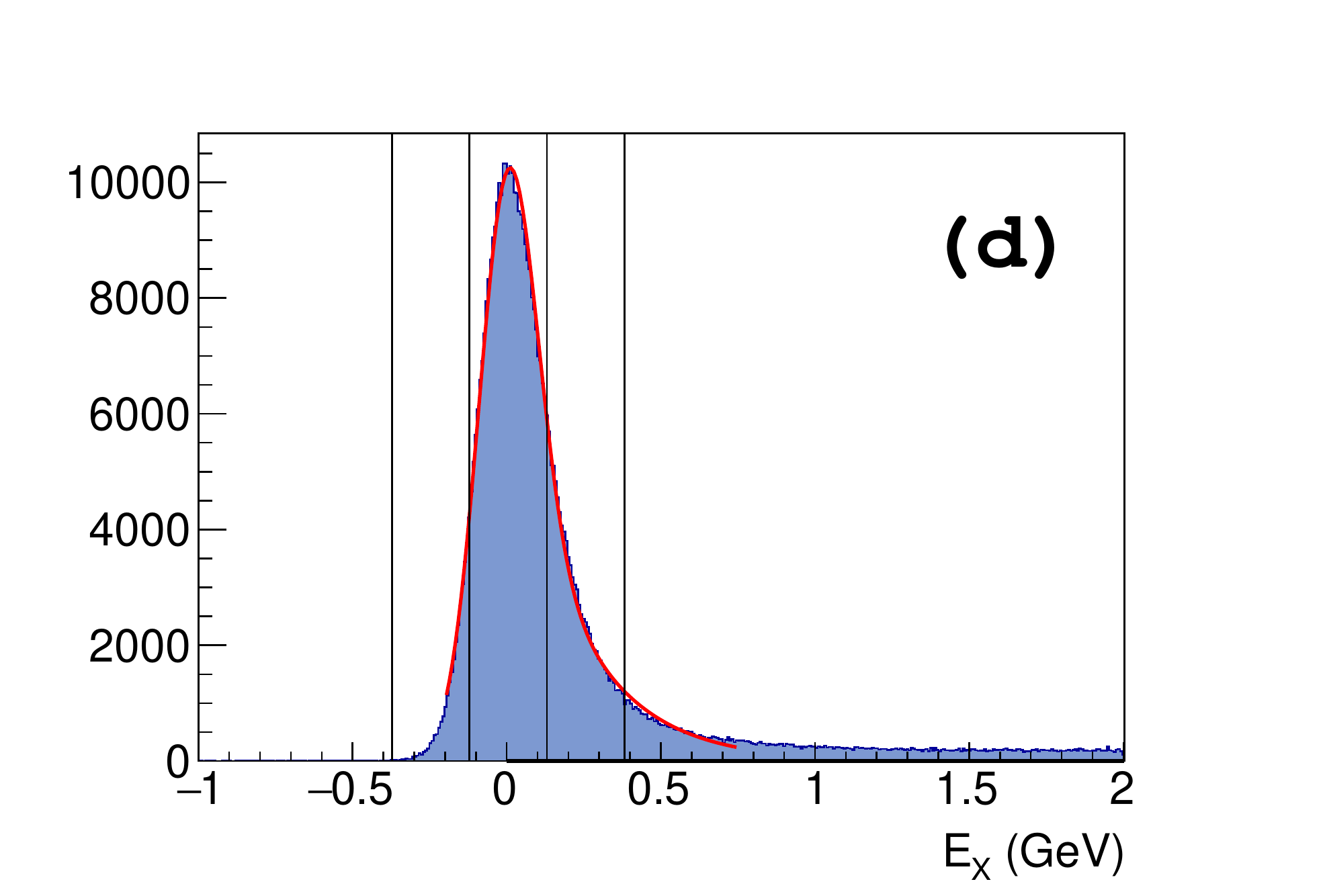}

\caption{The exclusivity distributions   after initial cuts. The upper  left  (a)  is $N(MM^{2}_{e+p\prime})$  for the data and the upper right (b) is $N(E_{X})$ for the data. The inner two lines are at $\pm$ 1 $\sigma$ and the outer two lines are at $\pm$ 3 $\sigma$ of the fitted distribution. The lower graphs (c and d respectively ) are the same distributions for Monte Carlo generated data. These distributions correspond to the majority of events where the final-state photon is detected in the IC. When the photon is detected in the EC, the distributions are somewhat wider and the cuts changed accordingly. The superimposed solid curves, in red,  in each graph are the results of fits to the distributions after initial cuts were applied on the other variables (see text). The missing mass squared was fit to a Gaussian and the missing energy to a skewed Gaussian. (Color online.)}
\label{fig:zcut2017_0}
\end{figure}

\begin{figure}[!ht]
\centering

\includegraphics[width=3.5in]{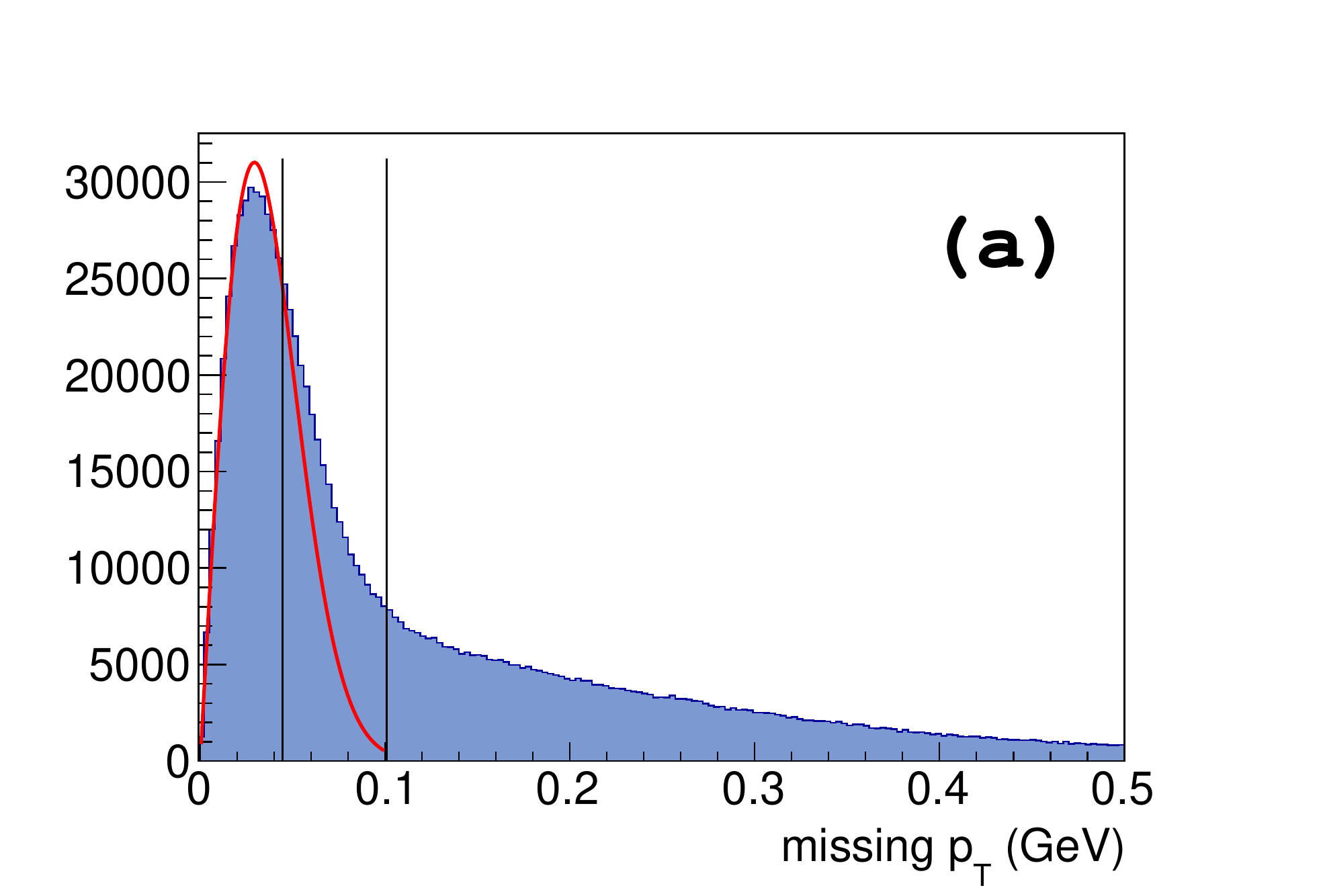}
\includegraphics[width=3.5in]{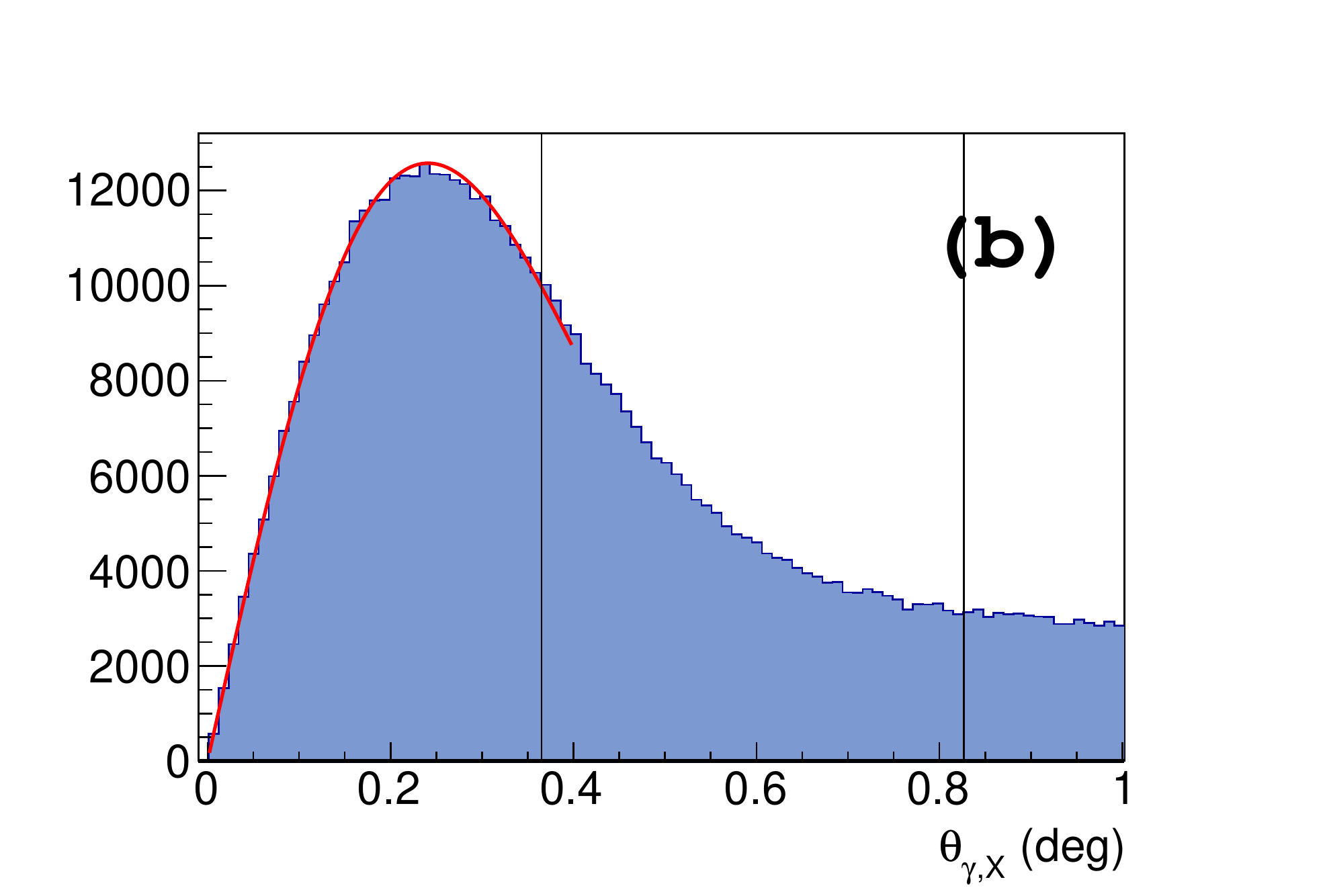}
\includegraphics[width=3.5in]{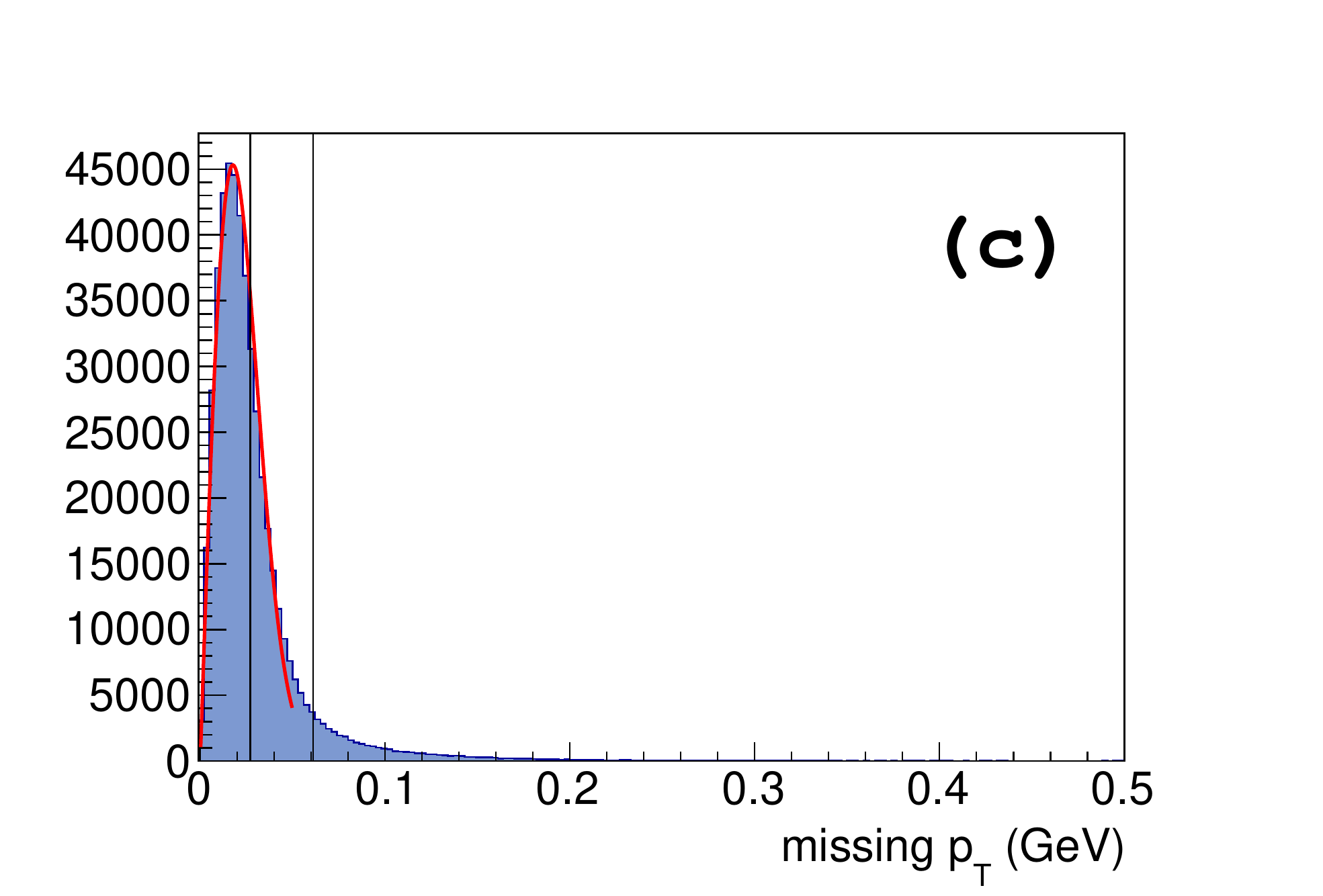}
\includegraphics[width=3.5in]{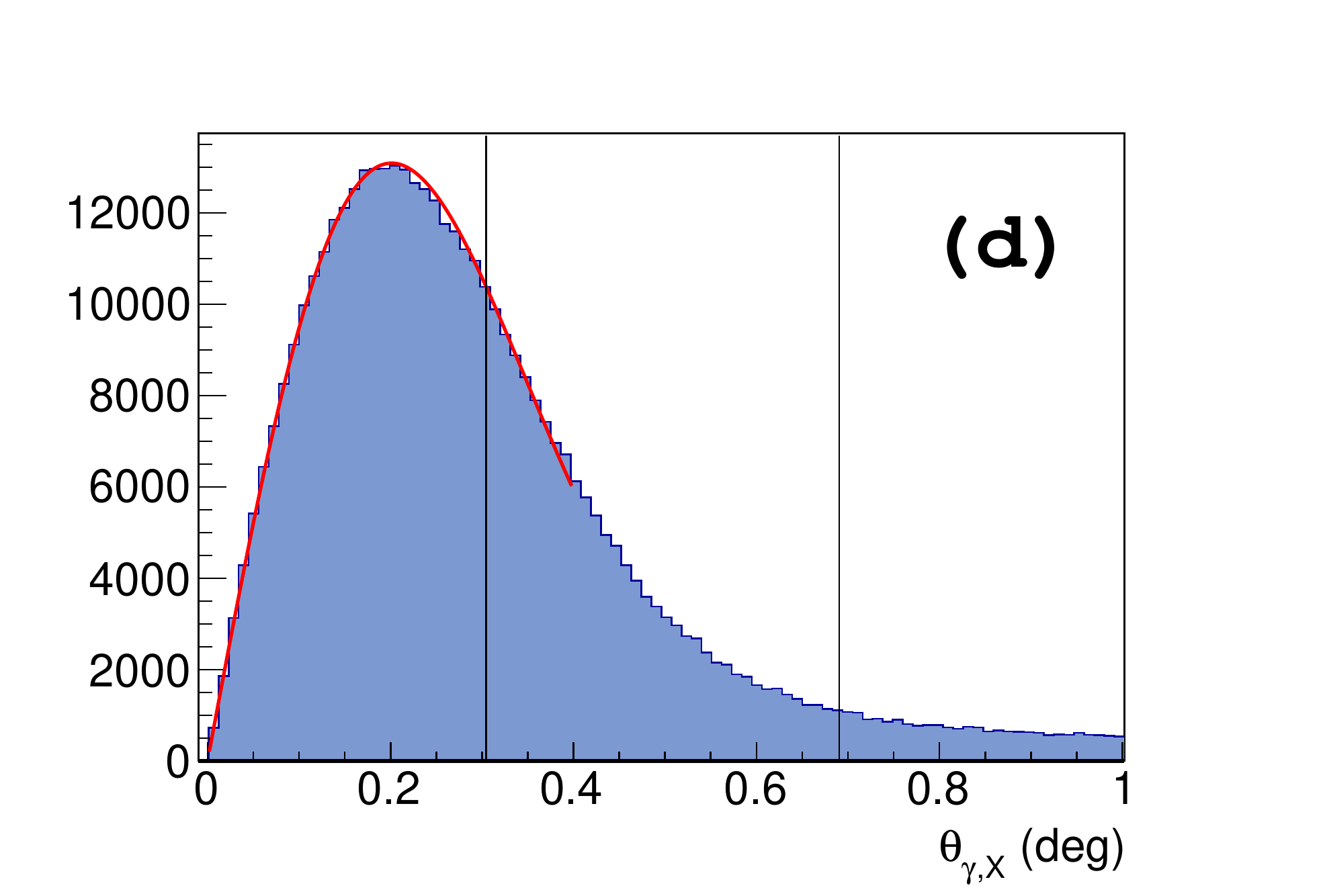}

\caption{The same as for Fig.{9}  
but for $N(p_{T})$, ( upper left, a) and $N(\theta_{\gamma,x})$ (upper right, b). The lower graphs, (c and d, respectively), are the same distributions for Monte Carlo generated data. The  distributions for $N(p_{T})$ and $N(\theta_{\gamma,x})$ were fit with empirical functions of the form  $A\sin(x\sigma)e^{(55\tan(x\mu) / \sigma)^2)}$, respectively  (Color online.)}
\label{fig:zcut2017_1}
\end{figure}

\begin{figure}[!ht]
\centering
\includegraphics[width=3.5in]{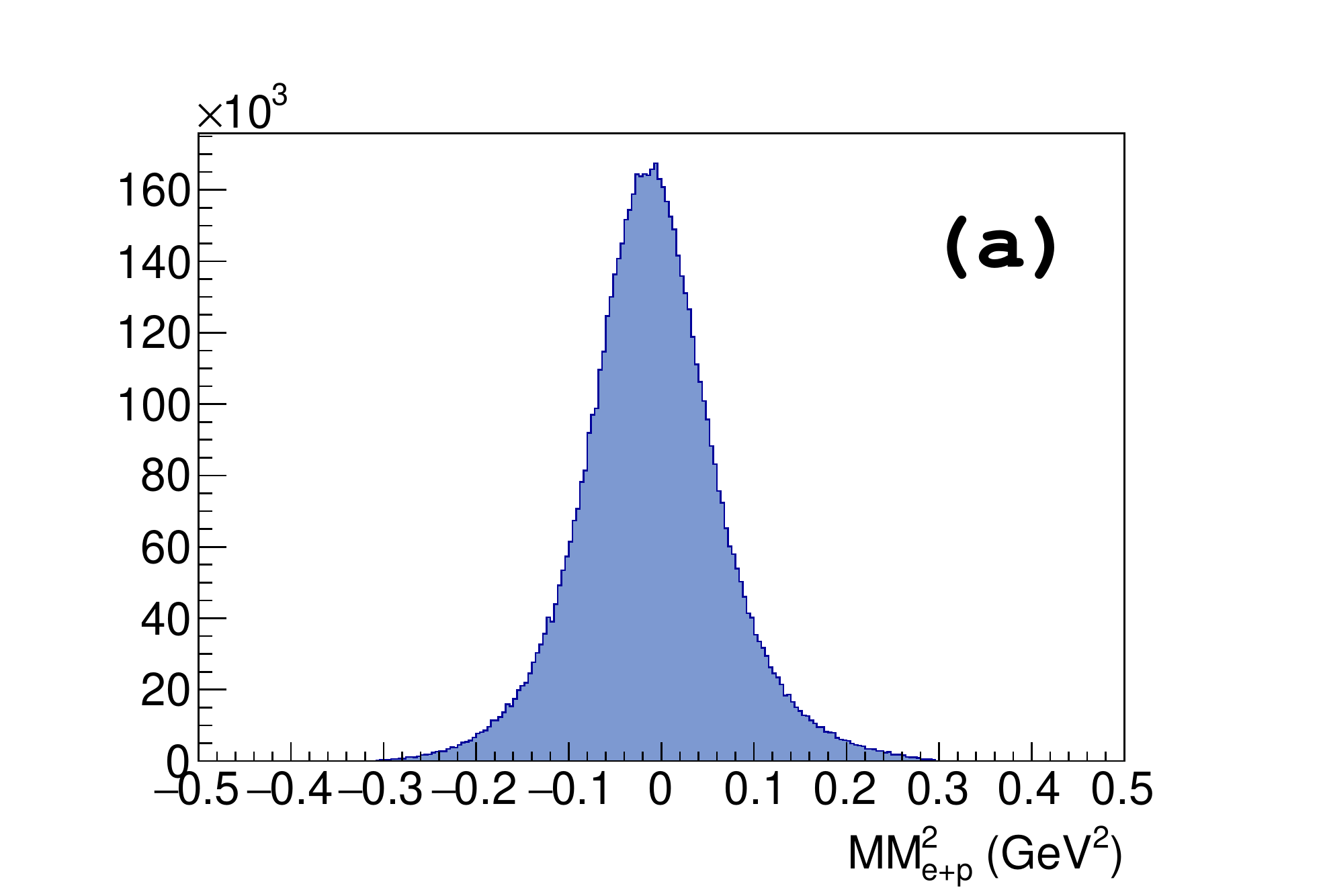}
\includegraphics[width=3.5in]{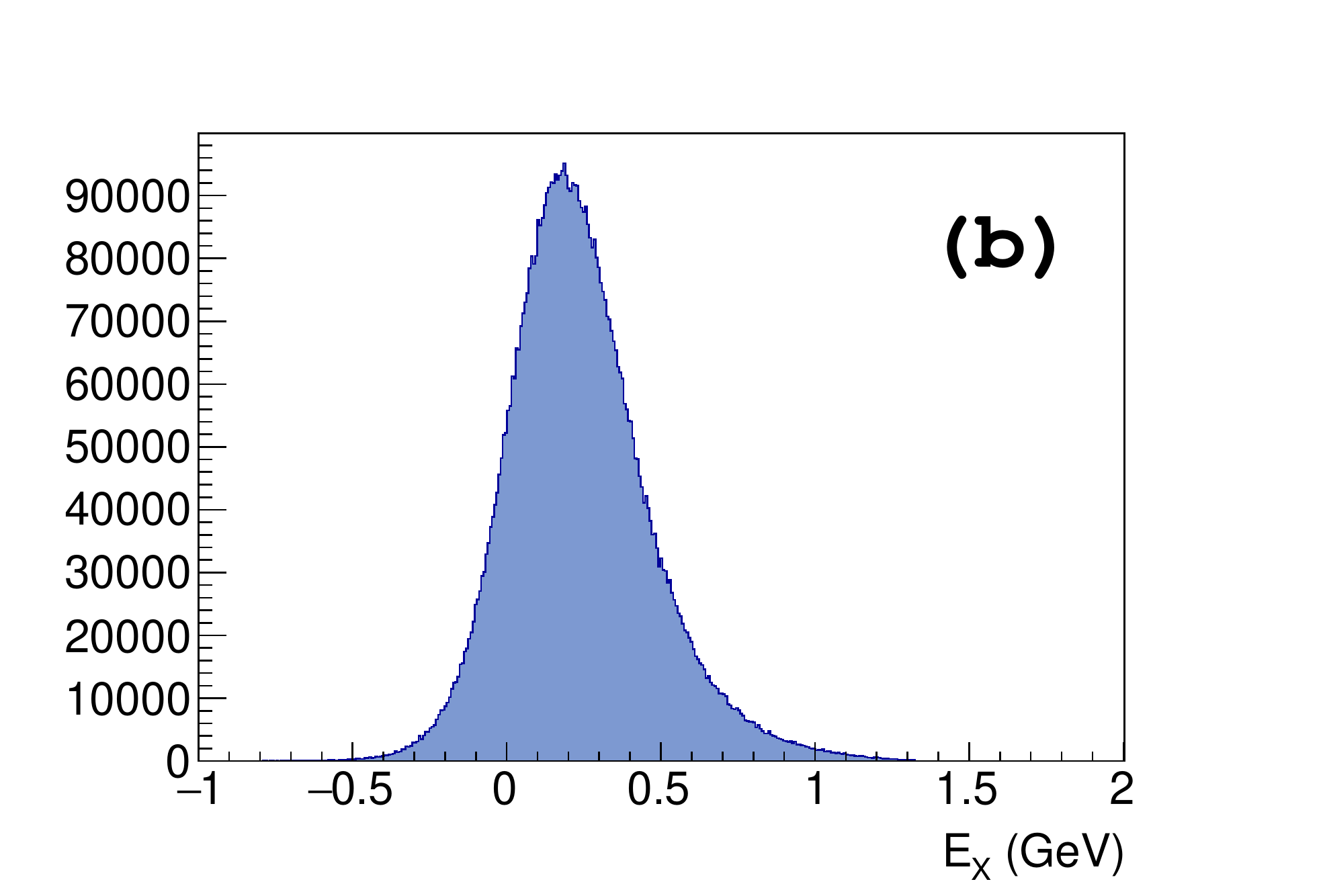}
\includegraphics[width=3.5in]{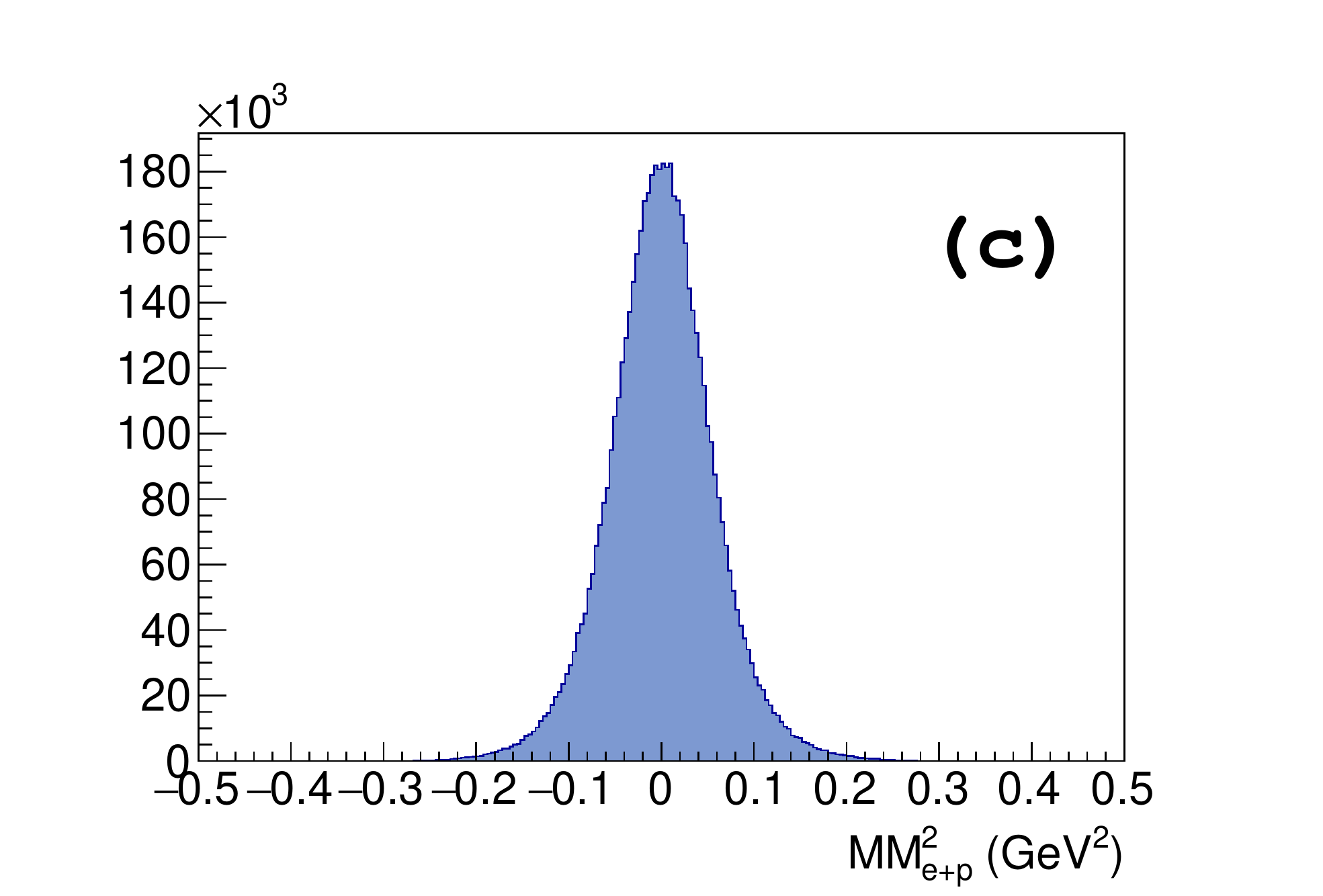}
\includegraphics[width=3.5in]{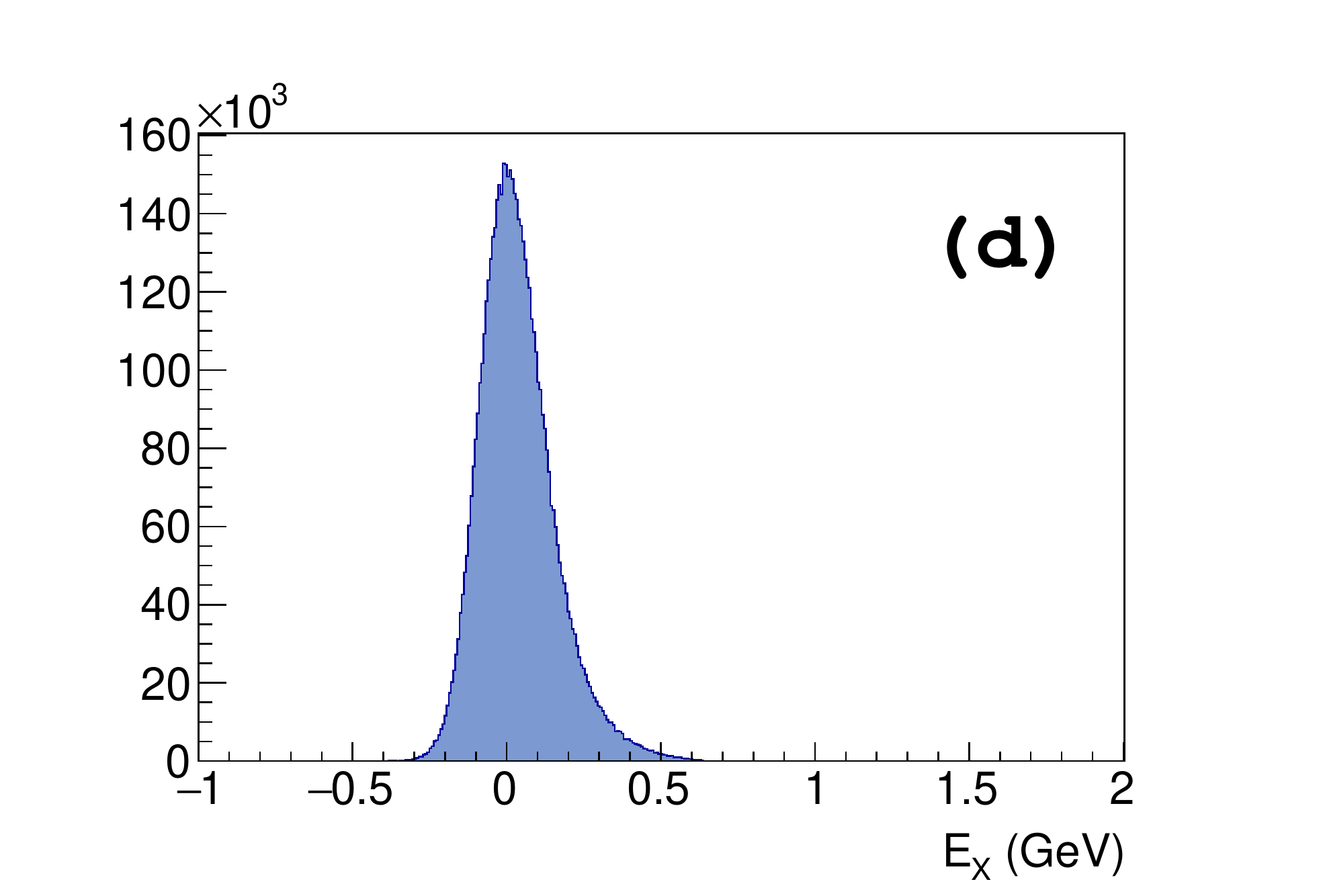}
\caption{The exclusivity distributions corresponding to Fig. 9 
   after all cuts. (Color online).The upper  left  is $N(MM^{2}_{e+p\prime})$  for the data and the upper right is $N(E_{X})$ for the data. The lower graphs, (c and d, respectively), are the same distributions for Monte Carlo generated data.}

\label{fig:zcut2017_2}
\end{figure}
\begin{figure}[!ht]
\centering
\includegraphics[width=3.5in]{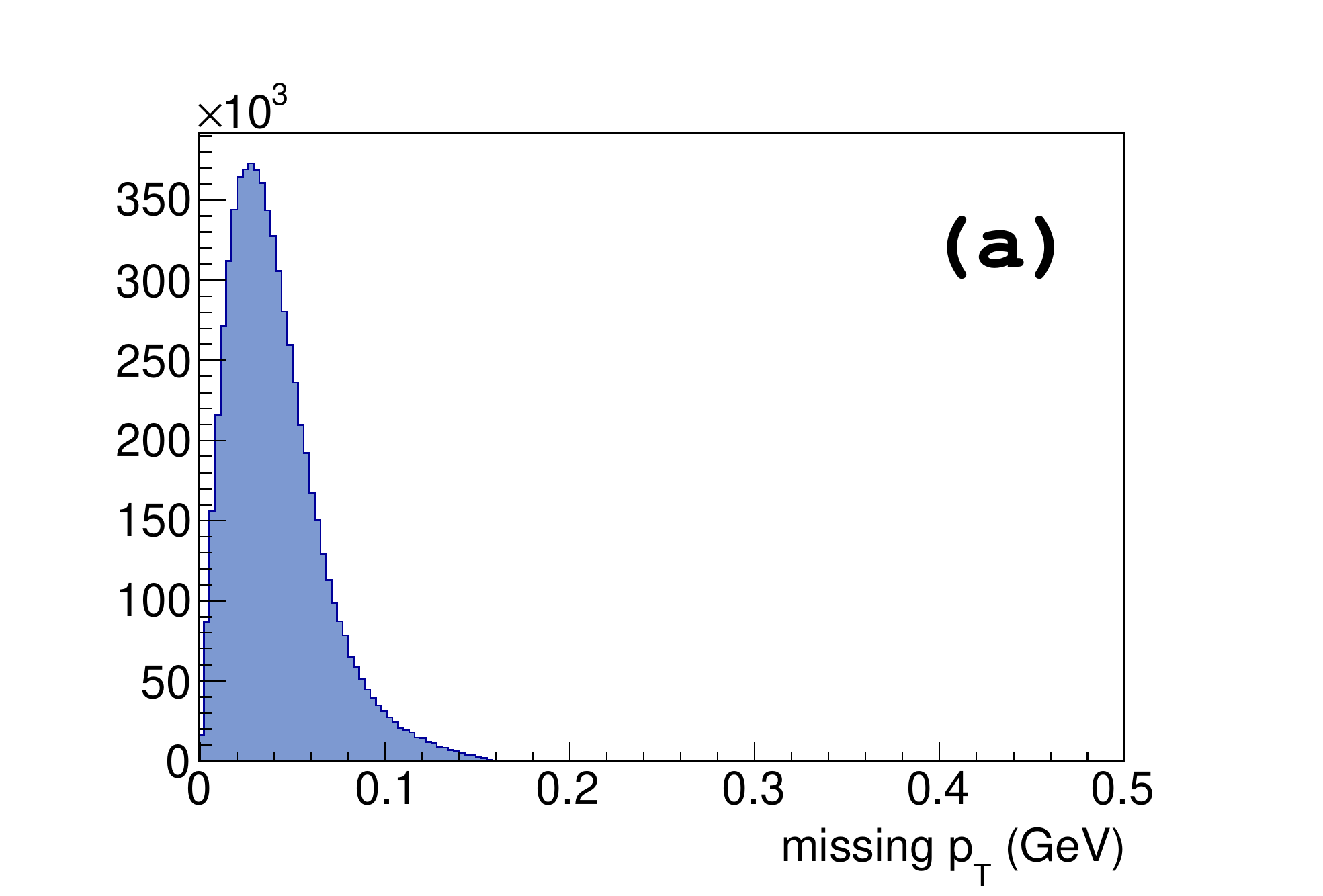}
\includegraphics[width=3.5in]{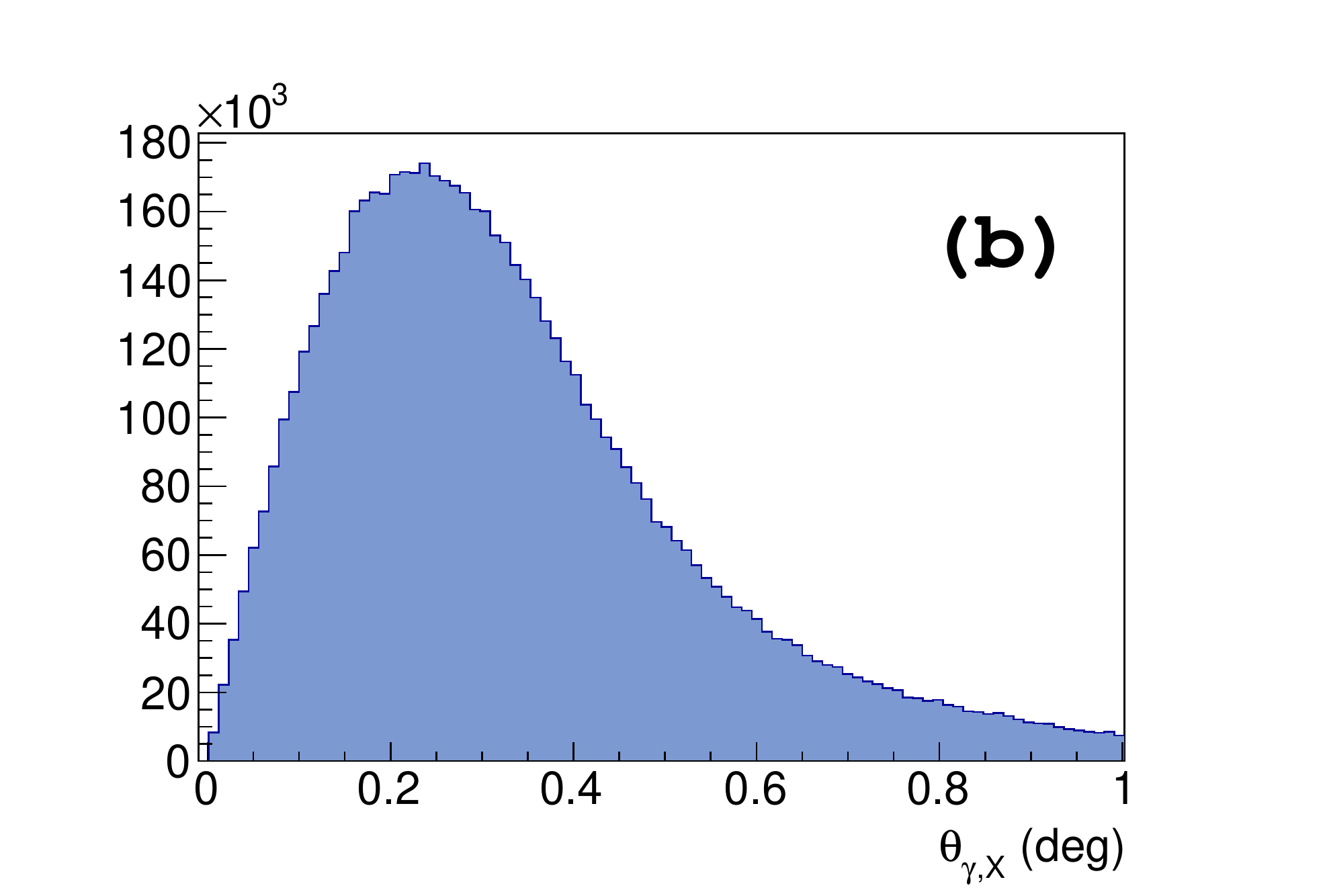}
\includegraphics[width=3.5in]{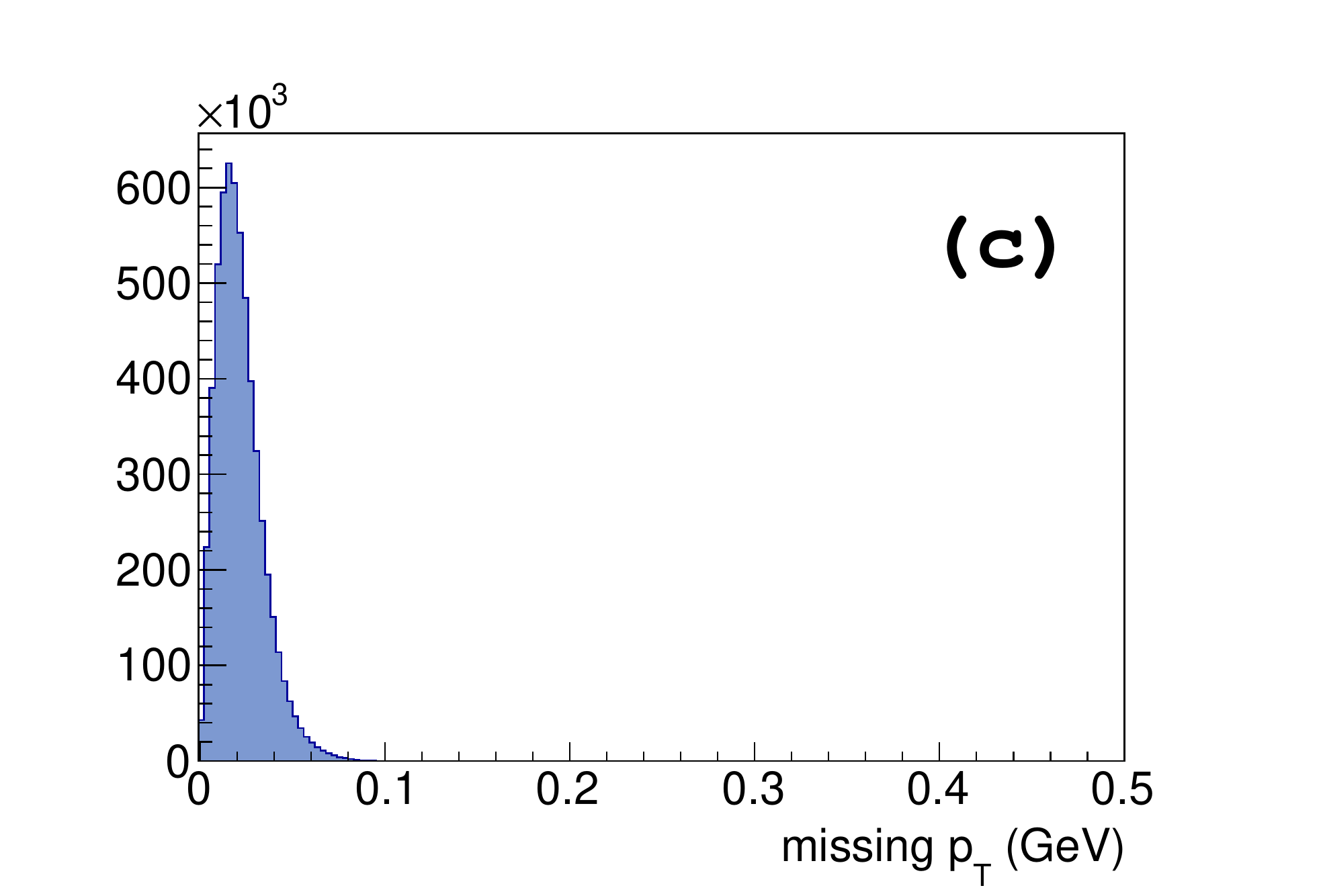}
\includegraphics[width=3.5in]{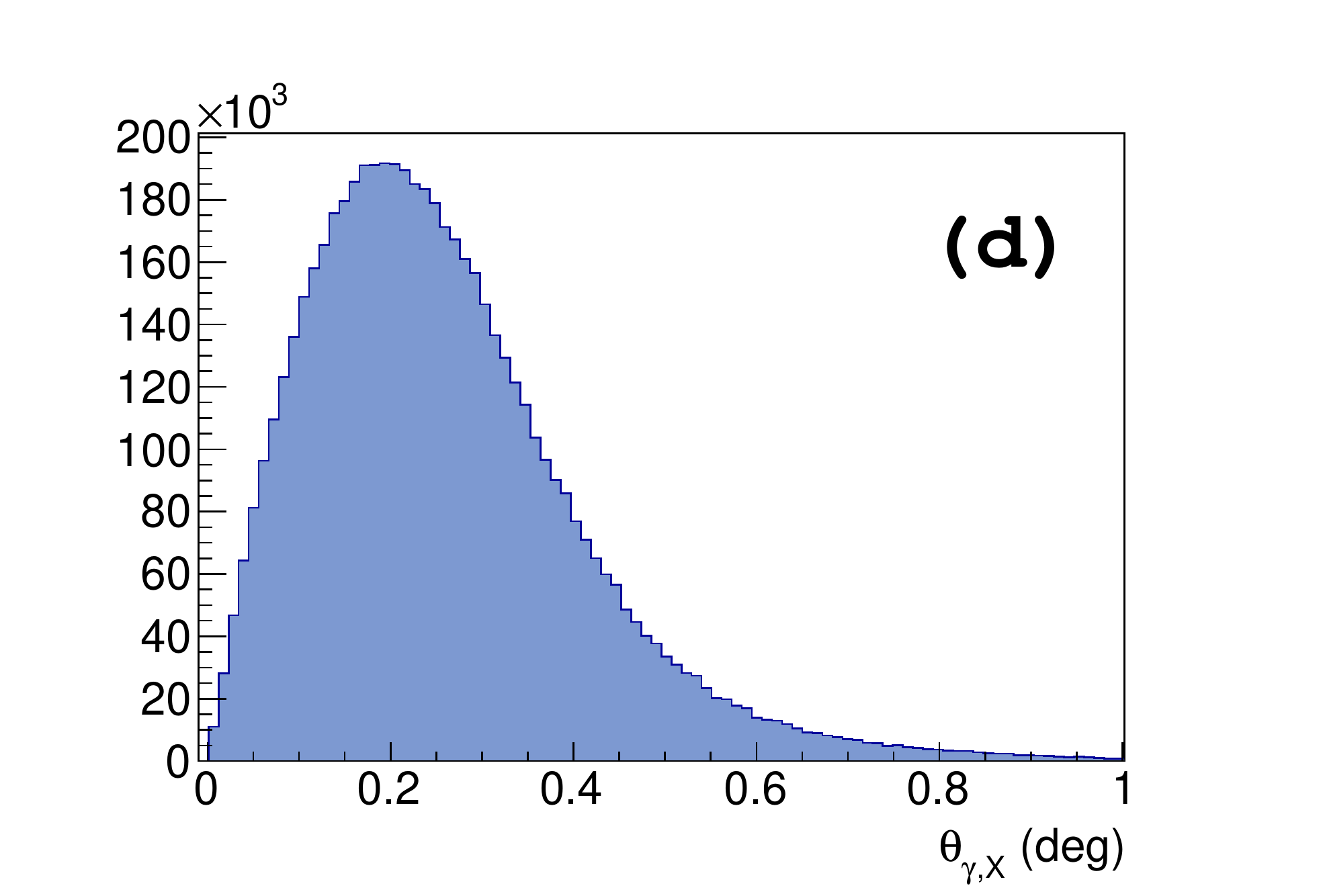}

\caption{Upper: The exclusivity distributions corresponding to Fig. 10 
after all cuts. Upper left (a) is $N(p_T)$.  Upper right (b) is $N(\theta_{\gamma,x})$. (Color online).The lower graphs, (c and d, respectively), are the same distributions for Monte Carlo generated data. }
\label{fig:zcut2017_3}
\end{figure}

\begin{figure}[!ht]
\includegraphics[width=7in]{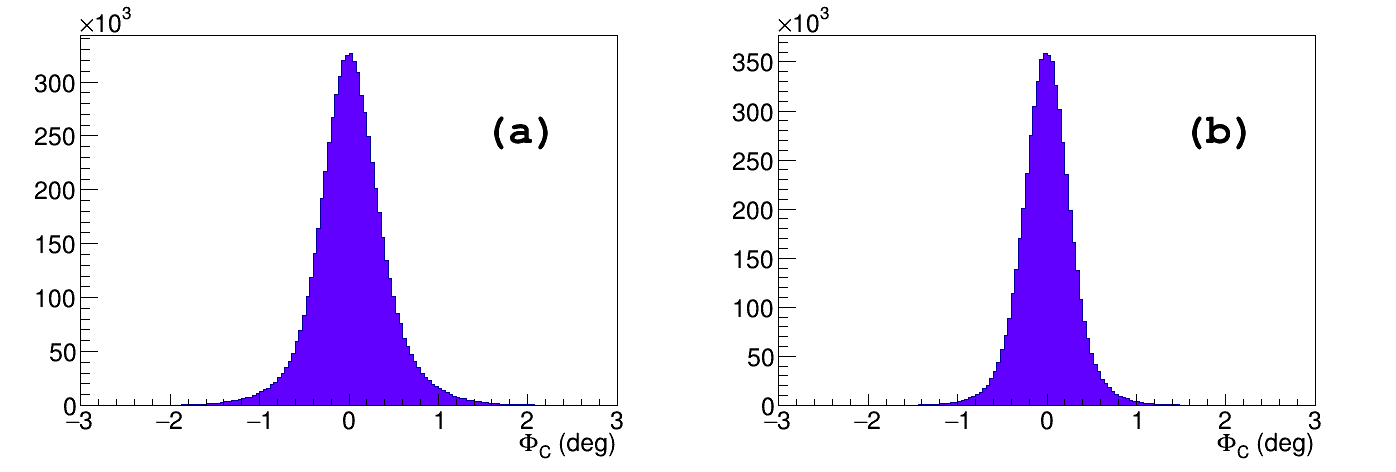}
\caption{The coplanarity  distribution $\phi_C$. On the left (a), for data after  cuts on $p_{T}$, $E_{X}$, $MM^{2}_{e'+p'}$, and $\theta_{\gamma,X}$, and on the (b)  right, for Monte Carlo generated data.   }
\label{fig.coplanarity}
\end{figure}

\FloatBarrier
\end{widetext}

\section{Monte Carlo Simulations}
\label{sec:mc}

\subsection{Definition of the Acceptance of CLAS}
The CLAS detector was not 100\% efficient in recording events due to acceptance losses from  gaps between the sectors associated with the torus coils and the inefficiencies associated with the edges of the different detector subsystems. In order to obtain a cross section, we had to estimate the fraction of events that were recorded by the detector to determine the acceptance. We define the acceptance $A$ as the fraction of events that were detected by CLAS. If the number of events detected by CLAS is $N_{\text{detected}}$, which we refer to as the yield, and the number of events which actually occurred is $N_{\text{actual}}$, which we refer to as the normalized yield, then the acceptance is defined by the relation:

\begin{equation}
A = \frac{N_{\text{detected}}}{N_{\text{actual}}}.
\end{equation}

We determined the acceptance from simulations of the experiment, including the knowledge of the detector geometry and properties using  the program  GSIM. 

The procedure  can be outlined in three steps.  First, an ensemble of events was generated. denoted $N_{generated}$. Second, 
the response of CLAS to each of these generated events was simulated, the results of which were reconstructed and processed by analysis software that reproduced the measured resolutions which are observed in CLAS. This output is denoted $N_{\text{reconstructed}}$. Third, the ratio between these two was calculated. This ratio is the acceptance:

\begin{equation}
A = \frac{N_{\text{detected}}}{N_{\text{actual}}} = \frac{N_{\text{reconstructed}}}{N_{\text{generated}}}.
\end{equation}

This acceptance $A$ is a function of the four variables: $Q^2,  x_B, t$, and $\phi$.
Since this analysis was performed bin-by-bin in the cross section variables, $A$ was accordingly  determined on a bin-by-bin basis.

\subsection{DVCS Generator}

An ensemble of DVCS plus BH events, $ep\rightarrow e'p'\gamma$, was generated  based on a parameterization of the theoretical model of Ref.~\cite{Belitsky:2001ns},
were fit to the data.  Additionally  radiative events were also generated, based on the calculation of Ref.~\cite{VFLMVV00}. These  included $ep\rightarrow e'p'\gamma + \gamma' + \gamma''$, where $\gamma'$ represents the radiated photon coming off the incoming electron leg, i.e. pre-radiation, and $\gamma''$ represents the radiated photon coming off the outgoing electron leg, i.e. post-radiation.

\subsection{GEANT3 Simulations (GSIM)}

Each of the detectors was simulated, specifying the geometry, placement, size, and material of each component. 
The EC was not perfectly simulated because of the complex showering that occured within it. To reduce the amount of computing time needed, any electrons or photons below a certain energy threshold were neglected, leading to  less  accuracy of the simulations. Also, the geometry of the CC was too complicated to  be implemented accurately in the simulations, so that the spectrum of the number of photoelectrons detected in the CC was not well reproduced. There were also resolution effects that were not taken into account in the simulations. In the cases of the SC and DC, the resolution in simulations was too narrow in comparison to the actual data. Some, but not all  differences between the simulated and detected responses were taken into account with a program called GSIM Post-Processer (GPP).

\subsection{GSIM Post-Processer (GPP)}

There were imperfections in the detector that where not included in GSIM. To simulate such effects, a post simulation processer called GPP was employed.  For example, DC sense wires that had  an efficiency of less than 1\% in the experiment were removed from the simulations. For those wires above 1\%  efficiency in the experiment, the corresponding wires in the simulation were modified to have the same efficiency~\cite{Baptiste,UNGAROLI}. Another function of GPP was to smear the DC and SC timing so that the missing mass resolutions of the Monte Carlo and the data matched.

\subsection{Background Merging}

During the experiment there was a possibility of M\o ller scattering, or other accidental events,  such as secondary scattering, cosmic rays, or other random events detected. Because of these background events, the efficiency of the reconstruction software was reduced in a non-trivial way. This background was not taken into account in GSIM. However, since this background was linearly related to the luminosity of the experiment~\cite{Baptiste}, it was therefore estimated in proportion to the luminosity of the experiment. Background events were then merged with the generated events in proportion to the background rate, which we measured from the experiment. It has been determined that the efficiency was reduced by about 6\%~\cite{Baptiste} due to the sources of accidental events.

\subsection{Comparison to Data}

Comparisons of kinematic distributions between data and the full Monte Carlo simulated events after all cuts showed  good agreement.  A comparison of each kinematic variable  is found in  Fig. \ref{fig:kineall}. The distribution of events in  $t$ vs. $x_B$ for data and Monte Carlo generated data are illustrated  Fig.~\ref{fig:kinetwo}.

\newpage
\begin{widetext}
\FloatBarrier
\begin{figure}[!ht]
\centering
\includegraphics[width=1.0\textwidth]{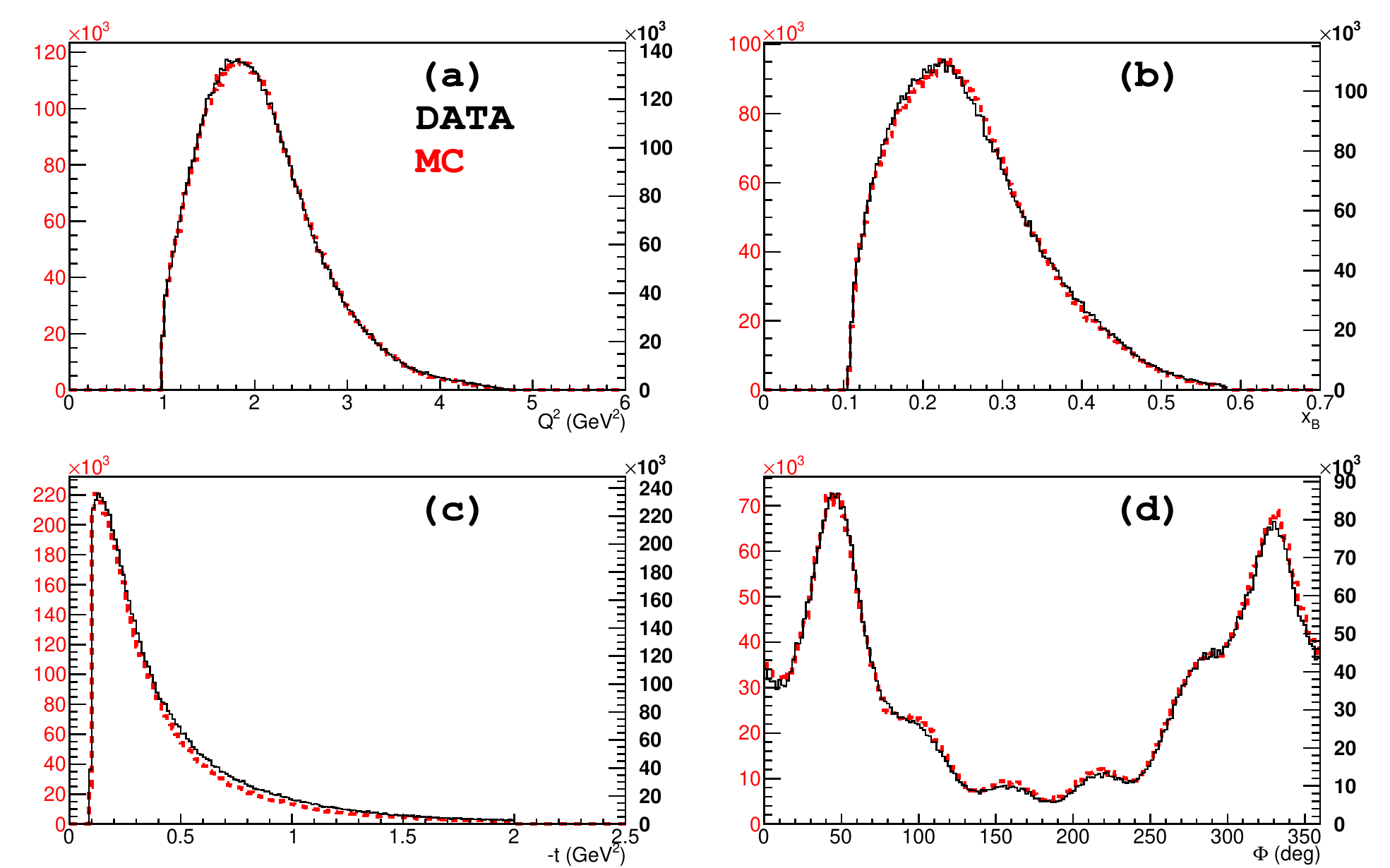}
\caption{A comparison of kinematic variable distributions for data versus Monte Carlo, upper left (a) vs $Q^2$, upper right (b) vs $x_B$,  lower left (c) vs $-t$, and  lower right (d) vs $\phi$.  The scales in red, on the left ordinate axes, correspond to Monte Carlo, shown as  red dashed curves. The scales in black, on the right ordinate axes,  correspond to data, shown as black solid curves (color online).
}
\label{fig:kineall}
\end{figure}

\begin{figure}[!ht]
\centering
\includegraphics[width=3.5in]{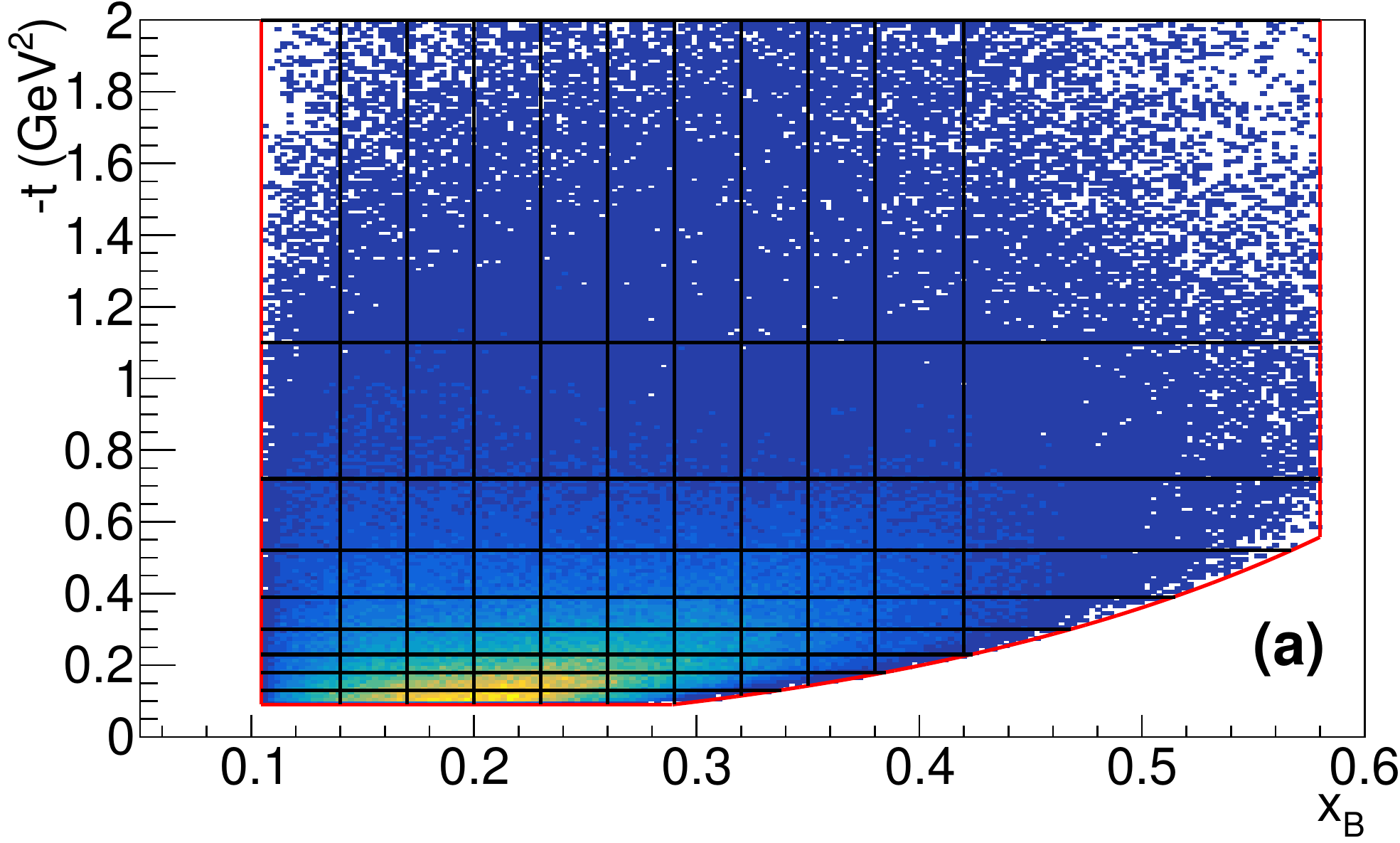}
\includegraphics[width=3.5in]{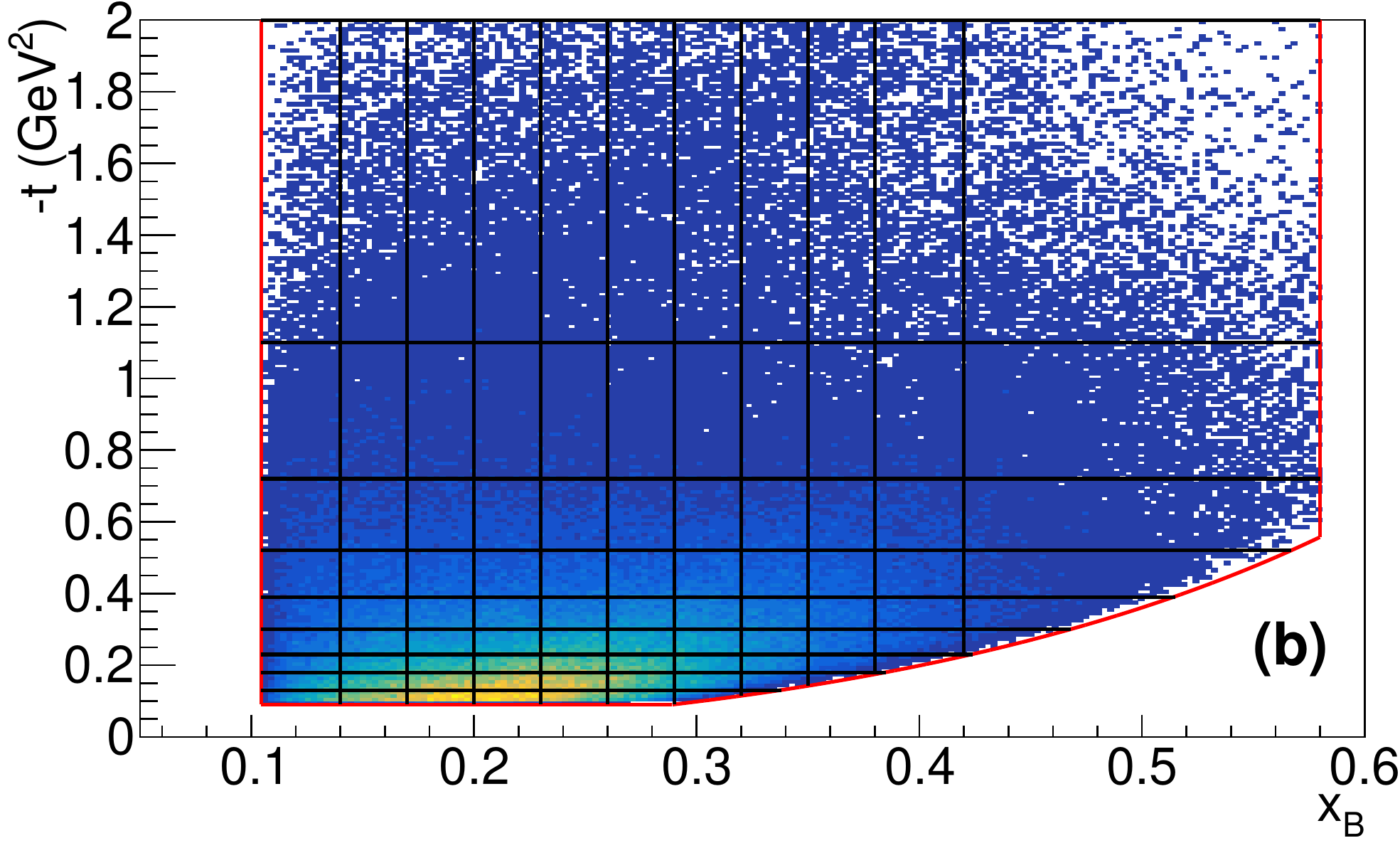}

\caption{The distribution of events in  $t$ vs. $x_B$ for experimental data (left, a) and Monte Carlo  generated data (right, b). There are 11 bins in $x_B$ and 9 bins in $-t$. (Color online.)}
\label{fig:kinetwo}
\end{figure}

\FloatBarrier
\newpage

\end{widetext}

\section{Neutral Pion Background Subtraction}
\label{sec:pi0}

\subsection{The Method for Estimating the Pion Contamination}
The exclusivity cuts that were placed on the $ep\rightarrow e'p'\gamma$ events were insufficient to remove all events
coming from the $ep\rightarrow e'p'\pi^0$ process. Therefore, an estimation of this $\pi^{0}$ contamination was made to correct the bin yields.

With regards to the $\pi^0$ decay, we considered two cases. First, the $\pi^0$ could decay such that both photons were detected. We refer to this  as $\pi^0 \rightarrow \gamma + \gamma$. Second the $\pi^0$ could decay such that only one photon was detected, and the other was missed either due to acceptance effects or the $150\ \text{MeV}$ photon detection threshold of the calorimeters. We refer to this as $\pi^0 \rightarrow \gamma + (\gamma)$, with parentheses indicating that the photon was not detected.

The actual number of $\pi^0$ decays with one photon detected, $N^{\gamma}_{\pi^0}$, could not be measured directly due to its indistinguishability with the 
$e'p'\gamma$ events. However, the number of $\pi^0$ decays with two photons detected, $N^{\gamma\gamma}_{\pi^0}$, could be measured.

In order to estimate the $\pi^{0}$ background, we note that both $N^{\gamma}_{\pi^0}$ and $N^{\gamma\gamma}_{\pi^0}$ can be used separately to determine the $\pi^0$ cross section:

\begin{equation}
\frac{\text{d}\sigma_{\pi^0}}{\text{d}\Omega} \propto \frac{N^{\gamma\gamma}_{\pi^0}}{A^{\gamma\gamma}_{\pi^0}},
\end{equation}

\begin{equation}
\frac{\text{d}\sigma_{\pi^0}}{\text{d}\Omega} \propto \frac{N^{\gamma}_{\pi^0}}{A^{\gamma}_{\pi^0}},
\end{equation}

\noindent where $A^{\gamma}_{\pi^0}$ and $A^{\gamma\gamma}_{\pi^0}$ correspond to the acceptances of each, as determined by Monte Carlo simulation. Since both are related to the $\pi^0$ cross section, one may write the following equation:

\begin{equation}
N^{\gamma}_{\pi^0} = N^{\gamma\gamma}_{\pi^0}\frac{A^{\gamma}_{\pi^0}}{A^{\gamma\gamma}_{\pi^0}}.
\end{equation}

Since the acceptances of each are just the ratios of the number of reconstructed particles to the particles generated, and only one generator is used in obtaining both acceptances, the equation can be further reduced:

\begin{equation}
N^{\gamma}_{\pi^0} = N^{\gamma\gamma}_{\pi^0}\frac{N^{\gamma}_{\pi^0, \text{rec}}}{N^{\gamma\gamma}_{\pi^0, \text{rec}}},
\end{equation}

\noindent  where the subscript ``rec'' corresponds to the reconstructed number of events in the Monte Carlo. From this relation, an estimation of the number of $\pi^0$ events with one photon detected was determined.

\vspace{0.25 in} 

\subsection{Computing $N^{\gamma\gamma}_{\pi^0}$ from Data}
\label{subsubsec:pi0data}

After selecting events with $ep\rightarrow e'p'(k  \gamma)$, where $k \geq 2$, every photon pair was looped over and we searched for pairs of photons with invariant mass near $m_{\pi^0}$. 
The photons selected are denoted as $\gamma_{1}$ and $\gamma_{2}$. For ease of reference, we define a four vector $p_{\pi^{0}} = p_{\gamma_{1}} + p_{\gamma_{2}}$, indicating that the combination is a $\pi^{0}$ candidate. In order to determine if they were truly pions, imposed  restrictions on certain quantities. First, for each of these combinations, the following quantities were calculated:

\begin{itemize}
\item $IM_{\gamma_{1}\gamma_{2}}$, the invariant mass of the photon pair of $ep\rightarrow e'p'\gamma_{1}\gamma_{2}$,
\item $MM^{2}_{e'p'X}$, the squared missing mass of the $ep\rightarrow e'p'X$ system,
\item $MM^{2}_{e'X\gamma_{1}\gamma_{2}}$, the squared missing mass of the $ep\rightarrow e'X\gamma_{1}\gamma_{2}$ system,
\item $E_{X}$, the missing energy of the $ep\rightarrow e'X\pi^{0}$ system.
\end{itemize}

The $\pi^{0}$'s were selected by requiring  $IM_{\gamma_{1}\gamma_{2}} $,
$MM^{2}_{e'p'X}$,
$MM^{2}_{e'X\gamma_{1}\gamma_{2}} $, and
$E_{X} $
 were within 3 standard deviations of the experimental resolution.

The resolution and calibration for the IC and EC were different. As a consequence, the distributions of these quantities appeared different based on which detector each photon entered. Each distribution was fit to a gaussian, or a gaussian plus background. The result was subtracted from the DVCS yield to remove the contamination.

The ratio of the $\pi_0 \rightarrow \gamma + \gamma$ yield to $\pi_0 \rightarrow \gamma + (\gamma)$ background, $\frac{A^{\gamma}_{\pi^0}}{A^{\gamma\gamma}_{\pi^0}}$, was found to depend on the kinematic bins in $Q^2, x_B, t$,  and $\phi$, smallest where the overall cross section is largest. For $\phi$ near $50^\circ$ or $310^\circ$, and $|t|$ near 0.1 GeV$^2$, where the yield is maximal, it was typically 1-2\%, while for $\phi$ near $180^\circ$, and  $|t|$ near 0.5 GeV$^2$, where the yield is minimal, it was on the order of 10\%.  

\section{Radiative Corrections}

The measured cross sections include the Born terms, in which we are interested, and radiative effects. Therefore, leading and next-to-leading order radiative corrections were calculated and used to obtain better measurements of the Born cross sections. The virtual photon corrections (vacuum and vertex) and real corrections (radiation) affect the measured cross sections in two distinct ways. The former interfere coherently, while the latter interfere incoherently with the DVCS Born terms. In the experiment, each of these diagrams is indistinguishable. We  took into account all virtual photon corrections, but radiative diagrams on the electron side only, recalling that the proton leg radiative diagrams are suppressed by the mass of the proton relative to the mass of the electron. In taking all leading order and next-to-leading order contributions into account, it is possible to calculate the radiative contributions to the measured cross sections. This has been worked out to leading order  in the soft photon approximation~\cite{Vanderhaeghen:2000ws}, and without the soft photon approximation up to next-to-leading order ~\cite{Akushevich:2012tw}. For this analysis, we used the corrections appearing in Ref.~\cite{Akushevich:2012tw}, which are estimated to have a systematic uncertainty of 3\%.

\begin{figure}[!ht]
\centering
\includegraphics[width=.4\textwidth]{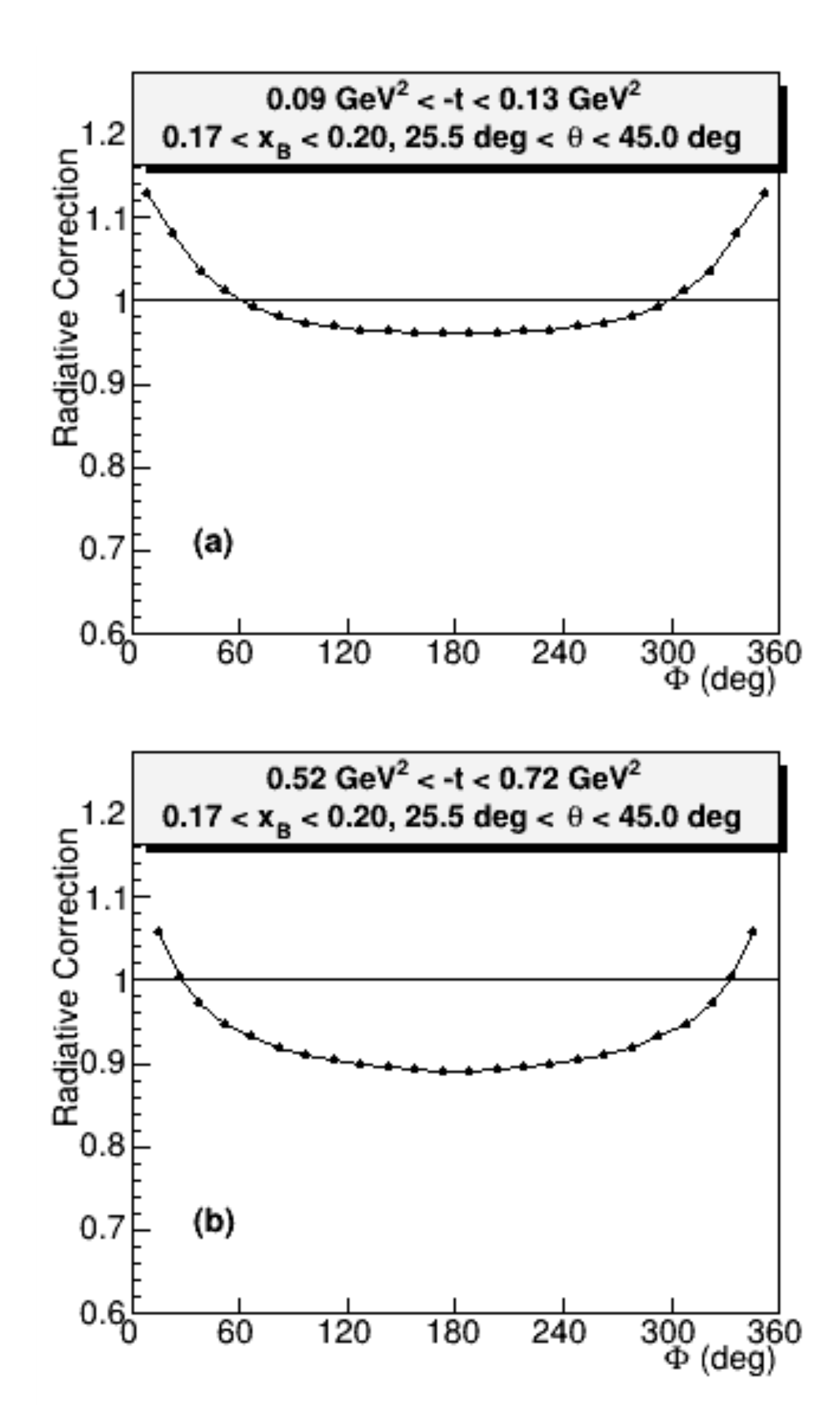}
\caption{Examples of  radiative corrections for $ep \to e'p'\gamma$ as functions of $\phi$ for two kinematic bins. }
\label{fig:radcor}
\end{figure}

The corrections, on average, increase the value of the measured cross section by 15\%. In general, the correction is larger near $\phi = 180^\circ$ and is smaller at large and small $\phi$. Examples of the radiative corrections as a function of $\phi$ for two kinematic bins are shown in Fig.~\ref{fig:radcor}. Since these corrections do not depend on the polarization of the incident electrons, they then apply to the polarized as well as unpolarized cross-section.

\section{Elastic Normalization}
After all the corrections were carried out there remained inefficiencies that were not accounted for, such as the inefficiency of the SC counters, reconstruction of tracks due to holes in the DCs and the SC counters which were not accounted for in GPP,  and accidental backgrounds.  To correct for these we followed procedures carried out for  previously published exclusive electroproduction cross section experiments performed at CLAS,  which utilized the same or similar experimental conditions (see Refs. \cite{Bedlinskiy:2012be,Bedlinskiy:2014tvi,Bedlinskiy:2017yxe,Jo:2015ema}).
This involved  a measurement of the elastic cross section as a function of $Q^{2}$ over a large range of the CLAS acceptance ~\cite{Baptiste}, and comparing to the cross section evaluation of Ref. ~\cite{Brash:2001qq}, which we denote as ``standard".

\begin{figure}[!ht]
\includegraphics[width=.4\textwidth]{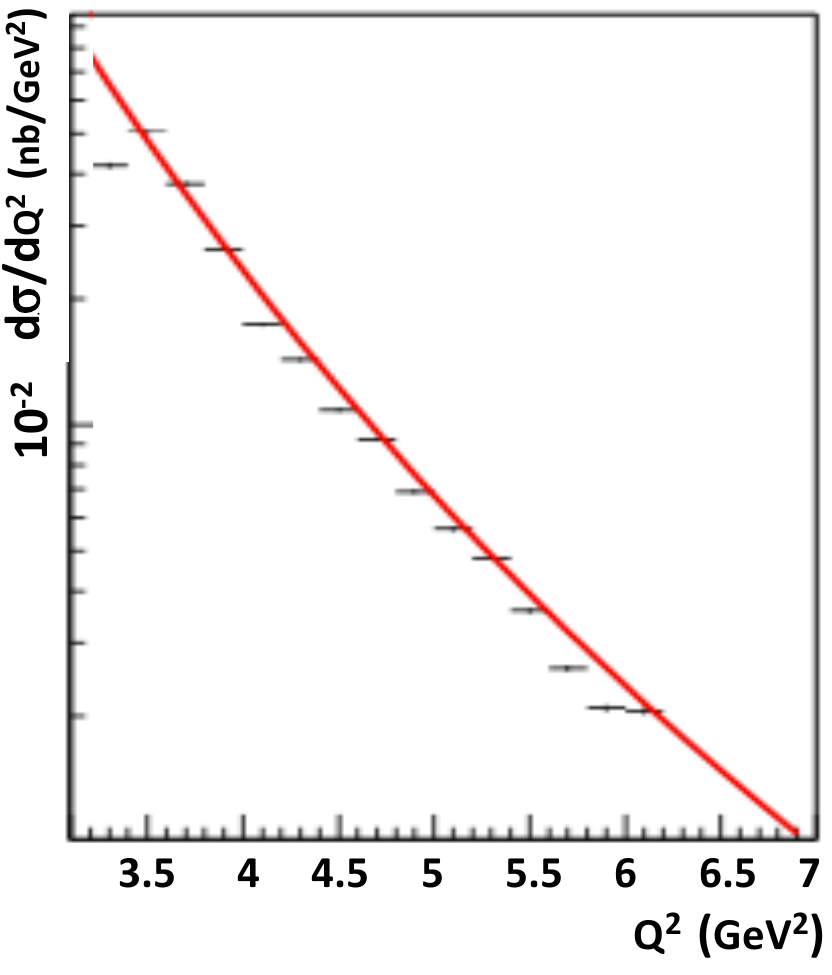}
\caption{The elastic cross section computed from e1-dvcs2 integrated over all sectors - black points. The cross section according to the parameterization of Ref.~\cite{Brash:2001qq} is displayed as a red line.(Color online.)}
\label{fig:elasticfinal}
\end{figure}

The elastic cross sections as a function of $Q^2$ obtained in CLAS and in Ref.~\cite{Brash:2001qq} are shown in Fig.~\ref{fig:elasticfinal}. The CLAS cross sections are somewhat lower than in Ref.~\cite{Brash:2001qq} for all kinematics. 

The integrated value of the ratio of  the cross sections obtained in CLAS and Ref.~\cite{Brash:2001qq}  over all kinematics,  $\epsilon = 0.926$,  was taken  as an overall normalization correction, where 
$$\epsilon = \frac{\text{d}\sigma}{\text{d}\Omega}_{\text{CLAS}} / \frac{\text{d}\sigma}{\text{d}\Omega}_{\text{standard}}.$$

Studies made using additional other reactions where the cross sections are well known, such as $\pi^0$ production in the resonance region, which are consistent with these normalization corrections over a wide range of kinematics covered by the present experiment. 

This correction comprises the largest single contribution to the systematic uncertainty in the extracted cross section, as noted in Section XA.

\section{Systematic Uncertainties}\label{systematics}
The major sources of systematic uncertainty are:

\begin{itemize}
\item Elastic normalization
\item Exclusivity cuts
\item Fiducial cuts
\item Radiative corrections
\item Beam polarization.
\end{itemize}

The estimated contributions from each of these sources is detailed in the following subsections.

\subsection{Elastic Normalization}

The systematic uncertainty of the elastic renormalization was estimated as the standard deviation from the mean as measured sector by sector:

\begin{equation}
\frac{\sigma}{\mu} = \frac{\sqrt{\frac{1}{6-1}\sum_{i=1}^{6}(\epsilon_{i} - \mu)}}{\mu} = \frac{0.037}{0.926} = 4.0\%,
\end{equation}
where $\mu$ is the average over all six sectors measured separately.

We conservatively assumed a global value of 5\% systematic uncertainty for the \emph{overall} global normalization, to account for any additional variations with kinematics of the renormalization factor  $\epsilon$.  Note, that this uncertainty comprises more than 50\% of the difference between the final normalized and  unnormalized cross sections.

\subsection{Exclusivity Cuts}

The systematic effects associated with the exclusivity cuts were obtained by varying each of the exclusive variable cuts. The variables used were: $E_{X}$, $MM^{2}_{e'+p'}$, $\theta_{\gamma,X}$, and $p_{T}$, defined in Sec.~\ref{sec:var}. 

We varied the exclusivity cuts on each of these variables and studied the response of the cross section as a function of those cut variations. 

The variation consisted in recalculating the cross section by applying new exclusivity cuts in steps of $\sigma/4$, from 1 $\sigma$ to 5 $\sigma$, where $\sigma$ corresponds to one standard deviation for each of the exclusivity variables. The systematic uncertainty was defined as half of the slope of the line tat was fit from 2.5-3.5 $\sigma$.

The uncertainty in the electron beam polarization was about 3\%, which was applied to the cross section differences.  

These systematic uncertainties were obtained for each kinematic bin. The average over all bins, was 5.5\%.

\subsection{Fiducial Cuts}

In order to have a measure of the systematic effect of the choice of fiducial cuts on the cross section, we  varied the fiducial cuts in much the same manner as the procedure for the exclusivity cuts. For this analysis, the cross sections were extracted using \emph{geometrical} fiducial cuts. In this study, we  maintained those cuts, and  applied new cuts, based on angles of tracks at the vertex which were based on reconstruction. Since this  procedure involves reconstruction of tracks through   the torus fields, it also tests  our knowledge  of the torus fields  which are used in simulations. 

The cuts we were momentum dependent, and were placed on all three final particles.  We then tightened these cuts in four steps of an eighth of a degree, azimuthally. This led to a total step of a half  degree. The step of half a degree was chosen because it represents the azimuthal resolution in the detector. The result is  4.2\%, which is  typically much less than the statistical uncertainty, and at most, on the order of the statistical uncertainty.

\subsection{Summary of Major Sources of Systematic Uncertainties}

Each of the systematic uncertainties,  averaged  over all kinematic bins, is presented in Table~\ref{tab:systematics}. These major sources of systematic uncertainty were assumed to be uncorrelated, so they were summed up in quadrature, leading to a total systematic uncertainty of 10.3\%. The overall systematic uncertainty is on the order of the statistical uncertainty. 

  \begin{table}[ht!]
      \begin{tabularx}{0.48\textwidth}{ l@{\hskip 1in}r@{}l }
        \multicolumn{3}{p{0.48\textwidth}}{}\\
        \hline
        \hline
        Source                & Err& or (\%)\\
        \hline
        Global Normalization  & 5&         \\
        Exclusivity Cuts      & 5&.5       \\
        Fiducial Cuts         & 4&.2       \\
        Radiative Corrections & 3&         \\
        Total Estimate        & 10&\%      \\
        \hline
        \hline
      \end{tabularx}
      \caption{Summary of systematic uncertainties. For the polarized cross sections, a systematic uncertainty of 3$\%$ on the beam polarization was added.}
      \label{tab:systematics}
  \end{table}

\section{Extraction of Unpolarized Cross Sections and Polarized Cross Section Differences}

The cross sections were obtained as in Eq.~\ref{eq:sig_ep_eppippim} for both positive and negative beam polarizations, and then combined to determine the unpolarized cross sections and polarized cross section differences. There were 189 $\phi$-dependent distributions,
corresponding to 189 ($x_B, Q^2, t$) bins, for each  beam polarization.

\subsection{Unpolarized Cross Sections}

Some examples of the unpolarized cross section are presented in Figs.~\ref{fig:cs09}, ~\ref{fig:cs49}, and \ref{fig:cs149}. The Bethe-Heitler process dominates at low and high $\phi$, and DVCS is more dominant in the central $\phi$ range. The unpolarized cross sections (see  Eq.\ref{Observables}), were determined from the data as follows:

\begin{equation}
\begin{aligned}
\frac{\text{d}^{4}\sigma_{\text{pol}}}{\text{d}Q^{2} \text{d}t \text{d}x_{B} \text{d}\phi} &= \frac{N }{\mathcal{L}_{\text{int}}A\Delta V F_{\text{rad}}\epsilon}.\\
\end{aligned}
\end{equation}

The experiment was deliberately carried out such that the integrated luminosities of each polarization were very nearly equal: $\mathcal{L}_{\text{int},+}=\mathcal{L}_{\text{int},-} = \mathcal{L}_{\text{int}} / 2$, so that $\mathcal{L}_{\text{int}}=\mathcal{L}_{\text{int},+}+\mathcal{L}_{\text{int},-} $. The number of events measured after the pion subtraction is $N = N_{+}+N_{-}$, viz.: 
$$\begin{aligned}
N_{+} = N^{e+p+\gamma}_{+} - N^{e+p+\pi^{0}(1\gamma)}_{+},\\ 
N_{-} = N^{e+p+\gamma}_{-} - N^{e+p+\pi^{0}(1\gamma)}_{-}.
\end{aligned}$$

\subsection{Polarized Cross Section Differences}
We  also extracted the polarized cross section differences. Some examples of the polarized cross section are presented in Figs.~\ref{fig:diff09},~\ref{fig:diff49}, and ~\ref{fig:diff149}. They were determined according to the following expression:

\begin{equation}
\begin{aligned}
\frac{\text{d}^{4}\sigma_{\text{pol}}}{\text{d}Q^{2} \text{d}t \text{d}x_{B} \text{d}\phi} &= \frac{1}{2}\left(\frac{\text{d}^{4}\sigma_{+}}{\text{d}Q^{2} \text{d}t \text{d}x_{B} \text{d}\phi} - \frac{\text{d}^{4}\sigma_{-}}{\text{d}Q^{2} \text{d}t \text{d}x_{B} \text{d}\phi}\right)\\
&=\frac{1}{2P}\left(\frac{N_{+}}{\mathcal{L}_{\text{int},+}} - \frac{N_{-}}{\mathcal{L}_{\text{int},-}}\right)\frac{1}{A\Delta V F_{\text{rad}}\epsilon},\\
\end{aligned}
\end{equation}

\noindent where $P$ corresponds to the beam polarization. For this experiment the polarization varied from about 0.83 to 0.87, and was taken overall to be at its average value of 0.853 with an uncertainty of 3\% of that value.  

Following Refs.~\cite{Belitsky01Asym, Belitsky:2001ns} the beam polarized cross section difference may be expressed in terms of ordinary form factors and CFFs at leading twist   as:

\begin{equation}
\Delta\sigma_{pol}\propto\sin(\phi)\text{Im}[F_1\mathcal{H}+\xi(F_1+F_2) \mathcal{\tilde H}-{{\Delta^2}\over{4m_p^2}}F_2\mathcal{E})],
\end{equation}

\noindent in which  $F_1$ and $F_2$ are the Fermi and Pauli  nucleon form factors, respectively,  and $\Delta$ the momentum transfer to the nucleon.  We note that the  pure BH and DVCS contributions have vanished in the polarized cross section. This is due to BH not being dependent on the polarization of the beam, and the fact that pure DVCS is dependent on beam spin only at the twist-three level. Again, due to the relatively small values of $x_{B}$ and $t$, the polarized cross sections are mainly sensitive to Im($\mathcal{H}$).

A table of all measured cross sections and  cross section differences  for all measured kinematic points can be obtained online from the CLAS database at \cite{clas-db} . 
\begin{widetext}
\FloatBarrier
\begin{figure}[!ht]
\centering

\includegraphics[width=1.\textwidth]{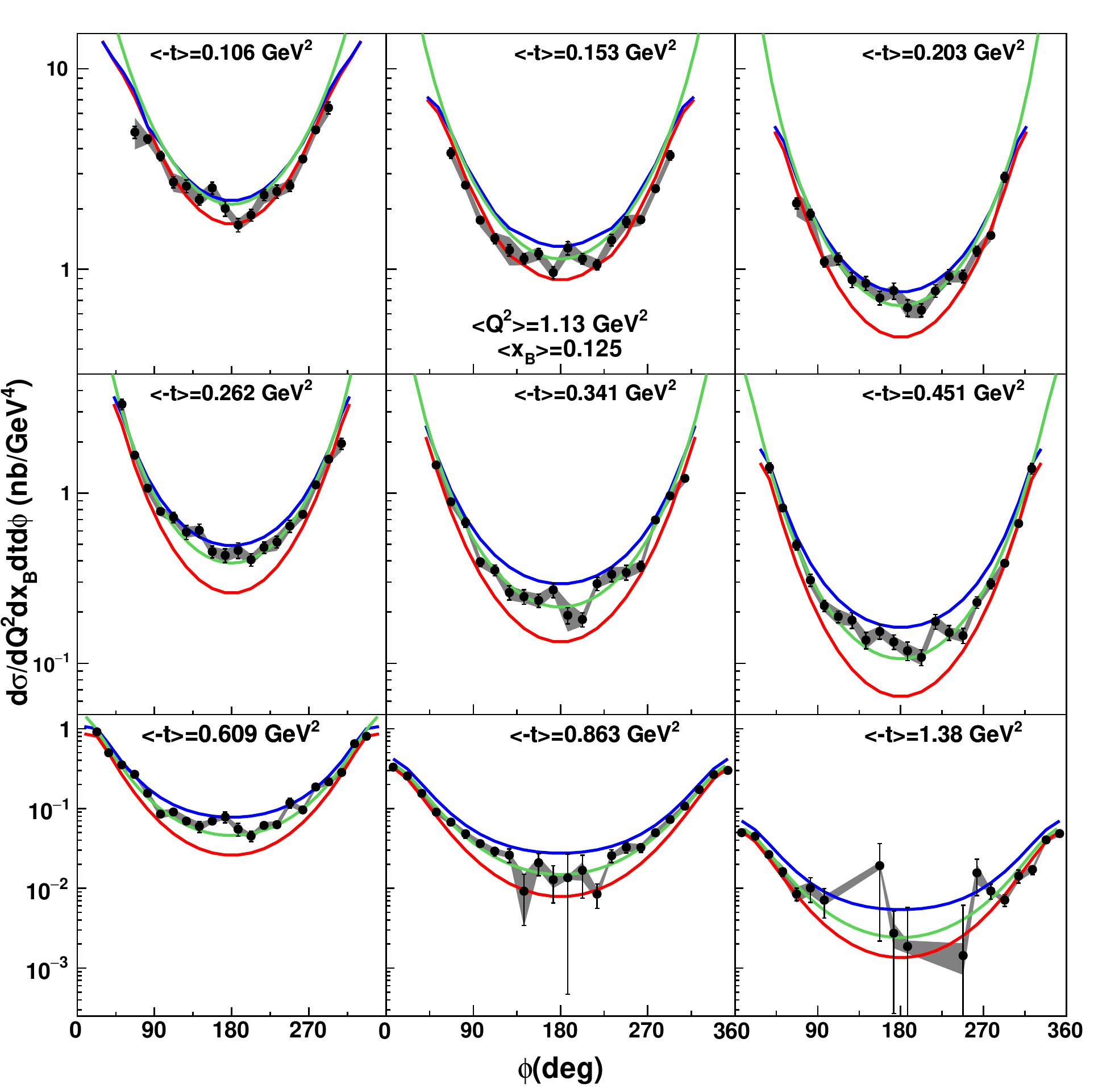}
\caption{The unpolarized cross section as a function of $\phi$ for the first kinematic bin at $\langle Q^2\rangle =1.13\  {\rm GeV^2 and } \langle x_B\rangle =0.125$ (see Fig.\ref{fig:KinematicalDomain} for bin definitions). The black points represent the results of the present experiment. The blue (upper) curves are the results of  the VGG model. The red (lower) curves are from BH contributions only. The green curves (beneath the VGG curves) are the result of the KMSC calculation. The bars on the data points are statistical uncertainties, and the grey bands represent the systematic uncertainties. (Color online.)
}
\label{fig:cs09}
\end{figure}

\begin{figure}[!ht]
\centering
\includegraphics[width=1.\textwidth]{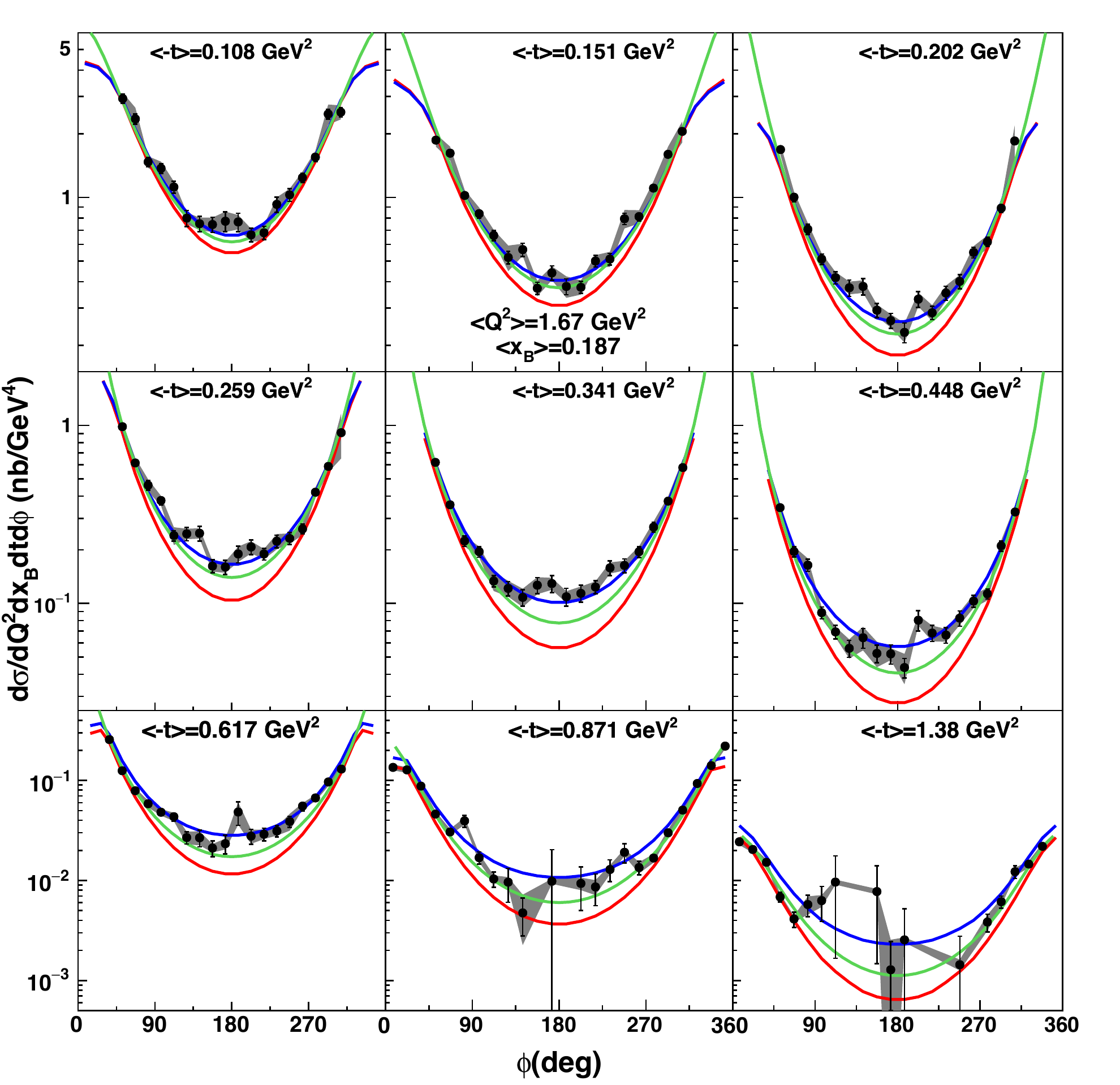}
\caption{The unpolarized cross section as a function of $\phi$ for the fifth bin at $\langle Q^2\rangle =1.67\  {\rm GeV^2 and } \langle x_B\rangle =0.187$. The black points represent the results of the present experiment. The blue (upper) curves are the results of  the VGG model. The red (lower) curves are from BH contributions only. The green curves (beneath the VGG curves) are the result of the KMSC calculation. The bars on the data points are statistical uncertainties, and the grey bands represent the systematic uncertainties. (Color online.)
}
\label{fig:cs49}
\end{figure}

\begin{figure}[!ht]
\centering

\includegraphics[width=1.\textwidth]{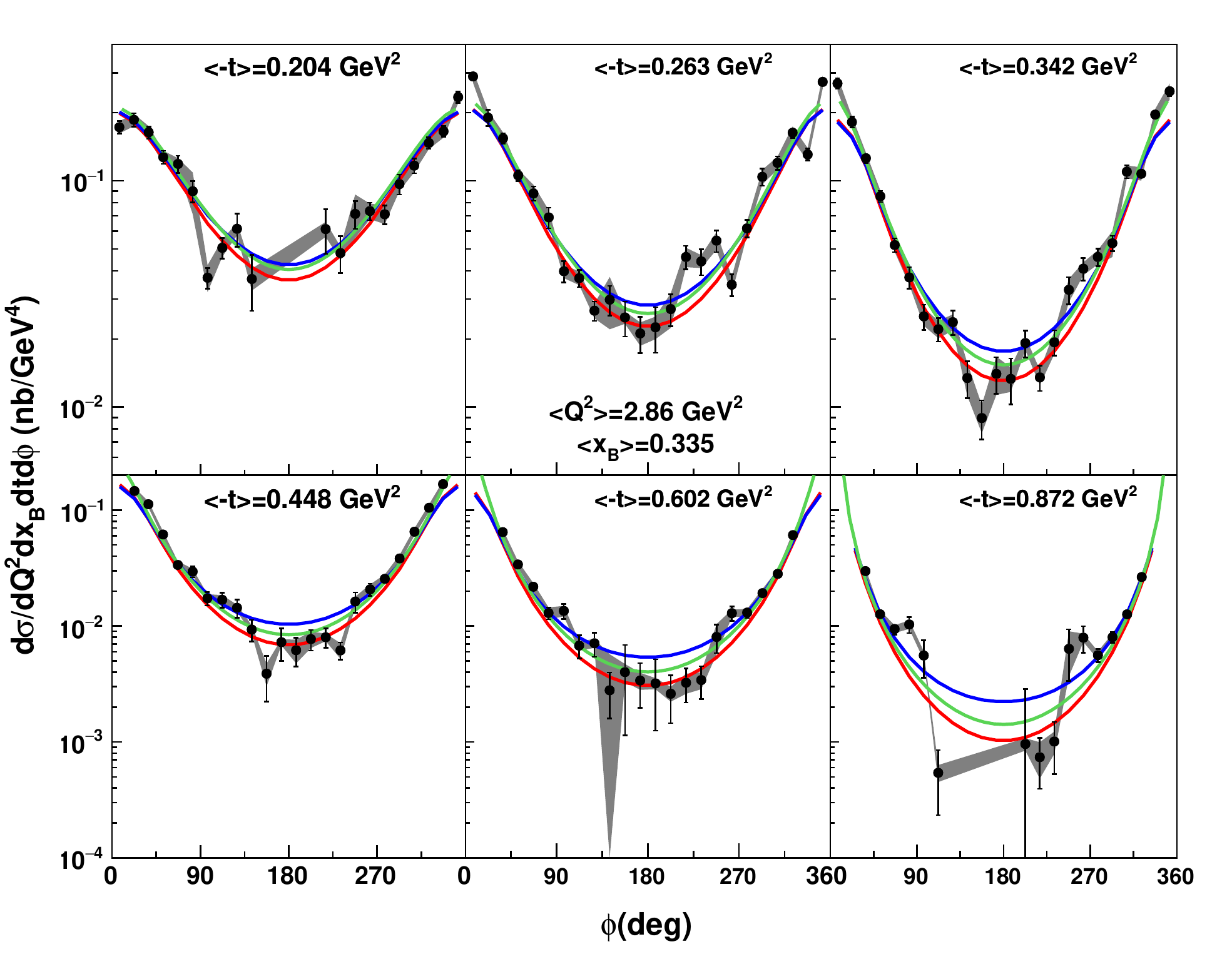}
\caption{The unpolarized cross section as a function of $\phi$ for the fifteenth bin at $\langle Q^2\rangle =2.86\  {\rm GeV^2 and } \langle x_B\rangle =0.335$. The black points represent the results of the present experiment. The blue (upper) curves are the results of  the VGG model. The red (lower) curves are from BH contributions only. The green curves (beneath the VGG curves) are the result of the KMSC calculation. The bars on the data points are statistical uncertainties, and the grey bands represent the systematic uncertainties. (Color online.)
}
\label{fig:cs149}

\end{figure}

\begin{figure}[!ht]
\centering

\includegraphics[width=1.\textwidth]{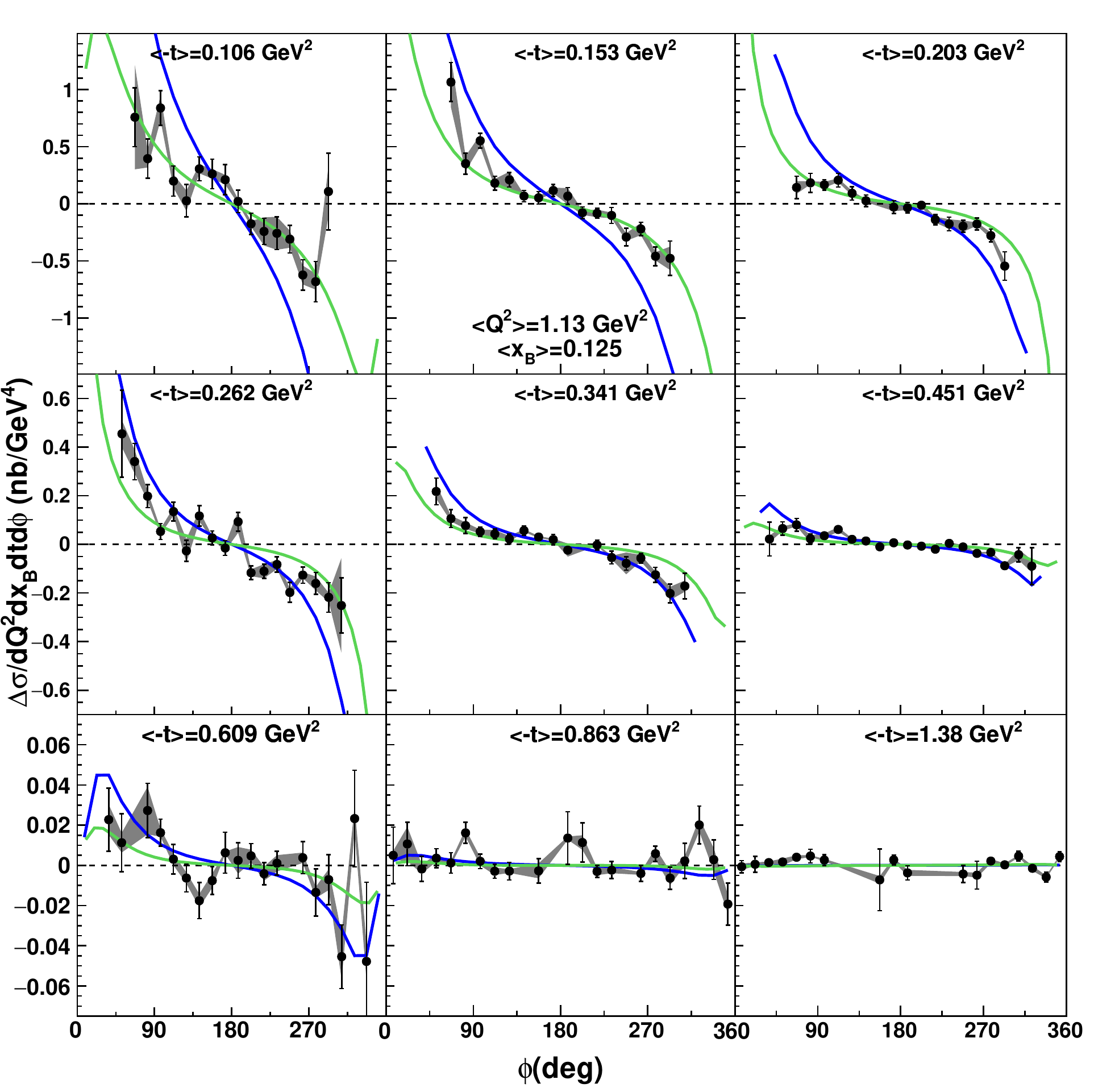}
\caption{The polarized cross section differences as a function of $\phi$ for the first bin kinematic bin at $\langle Q^2\rangle =1.13\  {\rm GeV^2 and } \langle x_B\rangle =0.125$. The black points represent the results of the present experiment. The blue curves, with generally larger asymmetry, are the results of  the VGG model, while the green curves, with generally lower asymmetry, are the results of KMSC calculation.The bars on the data points are statistical uncertainties, and the grey bands represent the systematic uncertainty. (Color online.)
}
\label{fig:diff09}
\end{figure}

\begin{figure}[!ht]
\centering

\includegraphics[width=1.\textwidth]{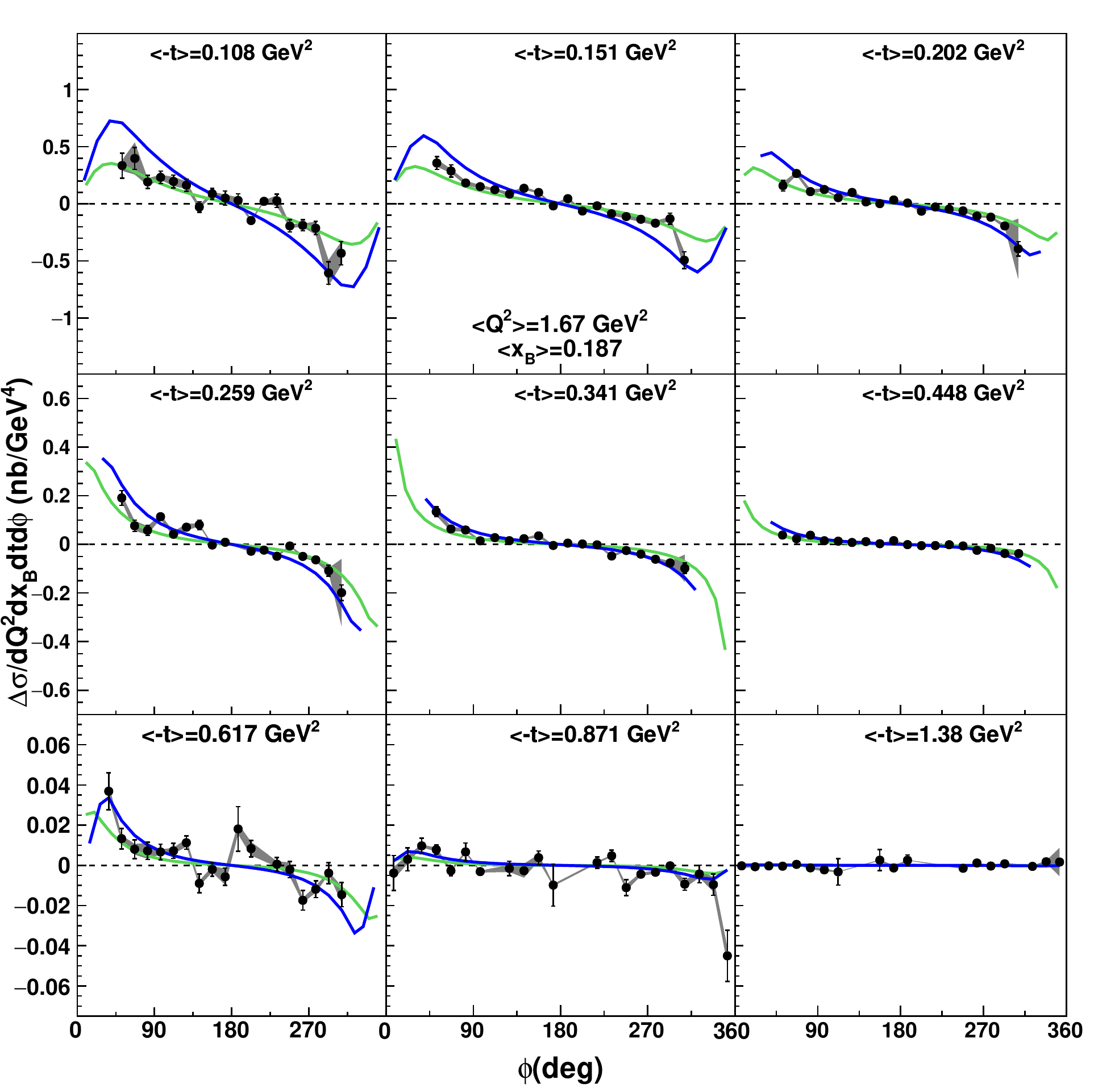}
\caption{The polarized cross section differences as a function of $\phi$ for the fifth bin at $\langle Q^2\rangle =1.67\  {\rm GeV^2 and } \langle x_B\rangle =0.187$. The black points represent the results of the present experiment. The blue curves, with generally larger asymmetry, are the results of  the VGG model, while the green curves, with generally lower asymmetry, are the results of KMSC calculation.The bars on the data points are statistical uncertainties, and the grey bands represent the systematic uncertainty. (Color online.)
}
\label{fig:diff49}
\end{figure}

\begin{figure}[!ht]
\centering

\includegraphics[width=1.\textwidth]{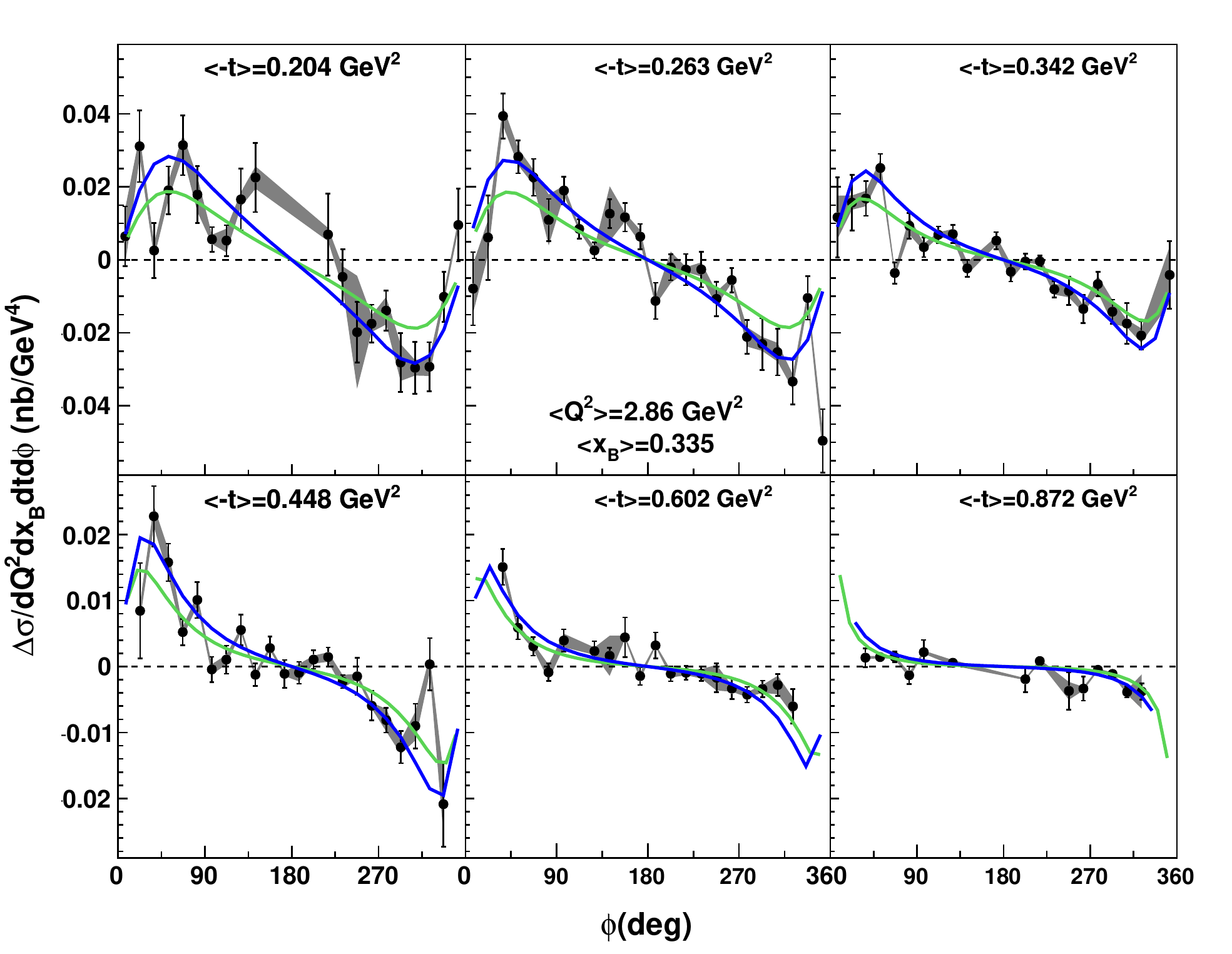}
\caption{The polarized cross section differences as a function of $\phi$ for the fifteenth bin at $\langle Q^2\rangle =2.86\  {\rm GeV^2 and } \langle x_B\rangle =0..335$. The black points represent the results of the present experiment. The blue curves, with generally larger asymmetry, are the results of  the VGG model, while the green curves, with generally lower asymmetry, are the results of KMSC calculation.The bars on the data points are statistical uncertainties, and the grey bands represent the systematic uncertainty. (Color online.)
}
\label{fig:diff149}

\end{figure}

\FloatBarrier
\end{widetext}

\section{Comparison with Previous CLAS Results}
The results presented in this paper originate from the second data taking run of the so-called e1-dvcs experiment. The results from the first run (e1-dvcs1) were already published \cite{FXGirod2008, Jo:2015ema}. In addition from being taken four years apart, the two runs differed by the beam energy (5.88 GeV presently vs 5.75 GeV previously), the exact positions of the target and of the  inner calorimeter with respect to CLAS, as well as the exact kinematics for each bin. 

In order to assess the compatibility of the two runs, a multiplicative factor was applied to the e1-dvcs1 cross sections \cite{Jo:2015ema} to account for the difference in beam energy. This factor originates from the known energy dependence of the Bethe-Heitler process as well as from the modeled, but lesser, dependence of the DVCS process. The factor  is $x_B, Q^2$  and  $\phi$ bin dependent, on average of the order of 4\%, and never exceeds 10\%. For a global comparison of the cross-section results, we calculated, for each of the 1907 four-dimensional bins, denoted $i$,  where the results overlap, cross section differences normalized by the combined uncertainties of the two runs:
\begin{equation}
\label{eq:compar}
\delta_i=\frac{\sigma 1_i-\sigma 2_i}{\sqrt{\Delta\sigma 1_i^2+\Delta\sigma2_i^2}},
\end{equation}
where for each run the uncertainties in the denominator are the quadratic sum of statistical and systematic uncertainties. 

\begin{figure}
\includegraphics[width=0.5\textwidth]{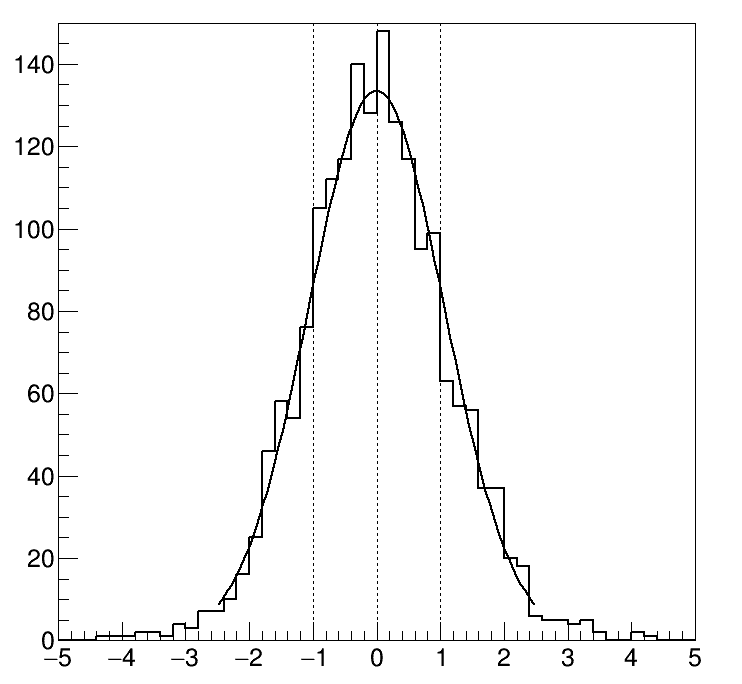}
\caption{Histogram of error-normalized cross section differences $\delta_i$ between e1-dvcs1 \cite{Jo:2015ema} and e1-dvcs2, as defined in Eq.~\ref{eq:compar}, with a fit to a Gaussian distribution (solid curve), which yields a mean of $0.001 \pm 0.028$ and a standard deviation of $1.06 \pm 0.024$.}
\label{fig:comp}
\end{figure}

Figure \ref{fig:comp} shows the results of the comparison between e1-dvcs1 and e1-dvcs2 in terms of these normalized differences. The two data sets are clearly consistent. The fact that this distribution is centered  nearly at 0 is a very good indication that the absolute normalization of both data sets is understood. A standard deviation of 1 indicates that the uncertainties are correctly evaluated. Likewise, the ratios of polarized to unpolarized cross sections were checked to be compatible with the published e1-dvcs1  beam spin asymmetries ~\cite{FXGirod2008}.

\section{Comparison with Model Calculations}
In this section, we compare the experimental cross sections with the theoretical calculations from the VGG~\cite{Vanderhaeghen:1998uc,Vanderhaeghen:1999xj,Goeke:2001tz,Guidal:2004nd, Guidal:2013rya} and KMSC~\cite{KMSC,KMS} models.
The former parametrizes the GPDs based on Radyushkin's double distributions ansatz with a few free parameters that are fitted to the nucleon form factor data. Only the GPD $H$ contribution is included and the parameter values are taken as: $b_v$ and $b_s$ (which control the $x$-$\xi$ correlation) are both equal to 1,  and $\alpha^\prime$ (which controls the $x$-$t$ dependence) is equal to 1.1.

The KMSC model also uses GPDs based on double distributions. The parameters were constrained by Deeply Virtual Meson Production data, nucleon form factors and parton distributions, the parameterization of the latter having been refined since the original calculation of Ref.~\cite{KMS}. All four chiral-even GPDs are included in the calculation and the DVCS amplitudes are calculated within the formalism of Ref.~\cite{BMJ} at leading-twist accuracy and next-to-leading order.

In Figs.~\ref{fig:cs09}-\ref{fig:diff149} we compare the results of the VGG and KMSC models to the unpolarized and the difference of beam-polarized cross sections from this work. For these calculations, the parameters of neither model have not been tuned to the current data. 

We selected three particular ($x_B$,$Q^2$) bins: (0.127,1.13), (0.186,1.67), (0.333,2.85), which exemplify the low, intermediate, and high ($x_B$,$Q^2$) domains spanned by the current experiment, respectively. For the three ($x_B$,$Q^2$) bins, we show the $\phi$-dependence of the cross sections for 8 or 9 $t$-bins. 
We recall that the $ep\to e'p'\gamma$ process is considered to be the coherent sum of the DVCS and BH contributions. 

In Figs.~\ref{fig:cs09}-\ref{fig:cs149}, the red curves show the contribution of the BH alone. The blue and green  curves show the sum of the BH and DVCS contributions. 
It is clear from their  $\phi$-dependence,  that the unpolarized cross sections, which peak at $\phi\approx 0^\circ$, are dominated by the BH contribution, especially near  $\phi$ = 0  and 180 deg.  Indeed, the BH cross section is maximal and quasi-singular when the outgoing real photon is collinear to the (incoming or outgoing) electron. This means that the photon is basically in the leptonic plane, i.e. $\phi\approx0^\circ$. The BH calculation is very well under control: the only non-QED inputs are the nucleon form factors, which, at the relatively low-$t$ values considered here, are well-known. Therefore, the differences between the data and the red curves correspond  to the DVCS contribution, which depends on the much less known GPDs.

For the beam-polarized cross sections, we observe that the  VGG and KMSC models generally reproduce the data, with VGG tending to somewhat overestimate and KMSC tending to underestimate the data.   We consider this quite satisfactory considering that parameters of the models have not been tuned.   
We conclude that the present data appear to be interpretable in terms of GPDs and have the potential to bring new constraints for CFFs or GPD extraction algorithms.
The KMSC model can be accessed on line from the PARTONS computing framework, Ref. \cite{KMSC} 
The spirit of this comparison is to show that the theoretical expectations are in fair agreement with the cross sections and cross section differences extracted in this work. The precise extraction of CFFs and GPDs requires a detailed and specialized fitting procedure, which is beyond the scope of this article. We refer the reader to references previously cited. 

\section{Summary and Conclusions}

The polarized and unpolarized cross sections for DVCS on the proton have been measured at a beam energy of 5.88 GeV with CLAS in a wide range of kinematics. Statistical and systematic uncertainties are each on the order of 10\%. These results will put constraints on GPD model parameters, and supplement past JLab data from Hall A and CLAS. We have presented a comparison of the current experiment with the VGG and KMSC models, and have compared the new results with the earlier DVCS experiment with CLAS (e1-dvcs1), which show reasonable compatibility. New experiments to further explore DVCS on the proton are currently active, and planned for the future at JLab both in Hall A and with CLAS12. All together these data will be used in fits for  extraction of GPDs.  The CLAS data, because of the large kinematic domain, will be very instrumental in constraining the dependences of the GPDs on the kinematical variables.

\section{Acknowledgements}

This work was supported in part by 
the Chilean Comisi\'on Nacional de Investigaci\'on Cient\'ifica y Tecnol\'ogica (CONICYT),
the Italian Istituto Nazionale di Fisica Nucleare,
the French Centre National de la Recherche Scientifique (CNRS), the French Commissariat \`a l'Energie Atomique (CEA), the French-American Cultural Exchange (FACE),
the United States National Science Foundation,
the Scottish Universities Physics Alliance (SUPA),
the United Kingdom's Science and Technology Facilities Council (STFC),
and the National Research Foundation of Korea  (NRF).

This material is based
upon work supported by the U.S. Department of Energy, Office of Science,
Office of Nuclear Physics under contract DE-AC05-06OR23177.

The Southeastern Universities Research Association (SURA) operates the
Thomas Jefferson National Accelerator Facility for the United States
Department of Energy under contract DE-AC05-84ER40150.

\begin{table}[ht]{\centerline {APPENDIX}}

Kinematics covered by the e1-dvcs2 experiment.
\begin{tabularx}{0.48\textwidth}{ c @{\hskip 0.5in}c @{\hskip 0.5in} c }
\multicolumn{3}{p{0.48\textwidth}}{\centering $x_B$ and $\theta_e$ binning}\\
\hline
\hline
$x_B$ and $\theta_e$ bin & $x_B$ & $\theta_e$ (deg.)\\
\hline
1  & 0.1-0.14  & 21-45 \\
2  & 0.14-0.17 & 21-25.5 \\
3  & 0.14-0.17 & 25.5-45 \\
4  & 0.17-0.2  & 21-25.5 \\
5  & 0.17-0.2  & 25.5-45 \\
6  & 0.2-0.23  & 21-27 \\
7  & 0.2-0.23  & 27-45 \\
8  & 0.23-0.26 & 21-27 \\
9  & 0.23-0.26 & 27-45 \\
10 & 0.26-0.29 & 21-27\\
11 & 0.26-0.29 & 27-45 \\ 
12 & 0.29-0.32 & 21-28 \\
13 & 0.29-0.32 & 28-45 \\
14 & 0.32-0.35 & 21-28 \\
15 & 0.32-0.35 & 28-45 \\
16 & 0.35-0.38 & 21-28 \\
17 & 0.35-0.38 & 28-45 \\
18 & 0.38-0.42 & 21-28 \\
19 & 0.38-0.42 & 28-45 \\
20 & 0.42-0.58 & 21-33 \\
21 & 0.42-0.58 & 33-45 \\
\hline
\hline
\\
\\
\end{tabularx} 

\begin{tabularx}{0.48\textwidth}{ c @{\hskip 1in}c}
\multicolumn{2}{p{0.48\textwidth}}{\centering  $t$ binning}\\
\hline
\hline
$t$ bin & $-t$  (GeV${}^2$)\\
\hline
1 & 0.09-0.13 \\
2 & 0.13-0.18 \\
3 & 0.18-0.23 \\
4 & 0.23-0.3 \\
5 & 0.3-0.39 \\
6 & 0.39-0.52 \\
7 & 0.52-0.72 \\
8 & 0.72-1.1 \\
9 & 1.1-2.0 \\
\hline
\hline
\\
\\
\end{tabularx}

\begin{tabularx}{0.48\textwidth}{ c @{\hskip 1in}c}
\multicolumn{2}{p{0.48\textwidth}}{\centering $\phi$ binning}\\
\hline
\hline
$\phi$ bin & $\phi$ \\
\hline
$1 \leq n \leq 24$ & $15 \times (n - 1)^\circ $ to $15 \times n^\circ $\\
\hline
\hline
\end{tabularx}
\end{table}

\bibliography{DVCS2-PRC}

\end{document}